%% file: LHSS.tex
\documentclass[twocolumns]{aa}
\usepackage{graphicx}
\def\amin{\ifmmode ^{\prime}\else$^{\prime}$\fi}
\def\asec{\ifmmode ^{\prime\prime}\else$^{\prime\prime}$\fi}
\def\h{$^{\rm h}$}
\def\m{$^{\rm m}$}
\def\s{$^{\rm s}$}
\def\etal{{et\,al.\,}}

\begin{document}
\title{ISOCAM observations in the Lockman Hole - I\thanks{Based on observations with ISO and Isaac
Newton Telescope (INT). ISO is an European Space Agency (ESA) project
with instruments funded by ESA Member States (especially the
P.I. countries: France, Germany, the Netherlands and the United
Kingdom) with the participation of the Institute of Space and
Astronautical Science (ISAS) and the National Aeronautics and Space
Administration (NASA). INT is operated on the island of La Palma by
the Isaac Newton Group in the Spanish Observatorio del Roque de los
Muchachos of the Instituto de Astrofisica de Canarias.}} \subtitle{The
14.3~$\mu$m shallow survey: data reduction, catalogue, and optical
identifications.\thanks{Tables 3 and 4 and FITS images are  available
in electronic form at the CDS via anonymous ftp to cdsarc.u-strasbg.fr
(130.79.128.5) or via http://cdsweb.u-strasbg.fr/cgi-bin/qcat?J/A+A/}}

   \author{D. Fadda
          \inst{1,2}
          \and
          C. Lari\inst{3}
	  \and
          G. Rodighiero\inst{4}
	  \and
	  A. Franceschini\inst{4}
	  \and
	  D. Elbaz\inst{5}
	  \and
          C. Cesarsky\inst{6}
	  \and
          I. Perez-Fournon\inst{2}
          }

   \offprints{D. Fadda}

   \institute{
	Spitzer Science Center, California Institute of Technology, 
	Mail Code 220-6, Pasadena, CA 91126, USA
	\email{fadda@ipac.caltech.edu}
         \and
	Instituto de Astrof\'\i{}sica de Canarias (IAC), Via Lactea S/N, E-38200 La Laguna, Spain
         \and
	Istituto di Radioastronomia del CNR (IRA), via Gobetti 101, I-40129 Bologna, Italy
         \and
	Dipartimento di Astronomia, Universit\`a di Padova, Vicolo dell'Osservatorio 5, I-35122 Padova, Italy
         \and
	CEA, DSM, DAPNIA, Service d'Astrophysique, F-91191 Gif-sur-Yvette Cedex, France 
         \and
	European Southern Observatory (ESO), Karl-Schwarzschild-Strasse, 2, 85748 Garching bei M\"unchen, Germany
}
   \date{}

   \date{Received date; accepted date}

   \abstract{ We present the image and catalogue of the 14.3~$\mu$m
shallow survey of 0.55 square degrees in the region of the Lockman
Hole (10\h52\m03\s $+$57$^o$21\amin46\asec, J2000) with the {\sl
Infrared Space Observatory} (ISO). The data have been analyzed with
the recent algorithm by Lari \etal (2001) conceived to exploit ISO
data in an optimal way, especially in the case of shallow surveys with
low redundancy.  Photometry has been accurately evaluated through
extensive simulations and also the absolute calibration has been
checked using a set of 21 stars detected at 14.3~$\mu$m, optical, and
near-IR bands.  On the basis of simulations, we evaluate that the
survey is 80\%, 50\%, and 20\% complete at 0.8, 0.6, and 0.45 mJy,
respectively. Below the 20\% completeness limit, fluxes are generally
overestimated since the sources are preferentially detected if their
positions correspond to positive oscillations of the noise. Moreover,
from a comparison with the deep survey, we estimate that only sources
brighter than 0.45~mJy are highly reliable. Only 5\% of these sources
do not have optical counterparts down to r'=25. Since none of the
Spitzer imaging bands cover the 14.3~$\mu$m wavelength range, this
data set will remain unique until the advent of the James Webb Space
Telescope.

 }

   \maketitle
%

\section{Introduction}

It is now widely accepted that a global vision of the universe can be
achieved only by complementing the ground-based optical observations with
satellite observations in wavelength domains unreachable from the
ground.  While violent phenomena like quasars and other active
galactic nuclei dominate the short wavelength extra-galactic emission
(X-ray and gamma), a large part of the star formation is obscured by
dust which reprocesses the UV-optical emission into infrared
radiation. So, the star formation activity in dusty regions can be
only observed indirectly through the emission of the dust in the
infrared or the synchrotron emission of electrons accelerated by
supernovae explosions in the radio.  The {\sl Infrared Astronomical
Satellite} (IRAS) in the local universe (Soifer, Houck, \& Neugebauer
1987) and subsequently the {\sl Cosmic Background Explorer} (COBE) with the
discovery of the cosmic infrared background (Puget \etal 1996, Fixsen
\etal 1998, Hauser \etal 1998) have dramatically shown that a large
part of the bolometric luminosity of the galaxies is emitted in the
infrared. In particular, the emission of the cosmic infrared
background which peaks at 140~$\mu$m represents more than half of the
overall cosmic background (Gispert, Lagache \& Puget 2000) while
approximately one-third of the bolometric luminosity of local galaxies
($z < 0.1$) is processed by dust into the infrared (Soifer \&
Neugebauer 1991).  This implies that the universe at $z>0.1$ is even
more active in the infrared than the local one shown by IRAS. 

For these reasons,  part of the guaranteed time of the mid-IR camera
ISOCAM (Cesarsky \etal 1996) was devoted to survey sky regions known
to have very low  neutral Hydrogen (HI) absorption to explore the deep
universe without any interference from our own galaxy. The primary
goal of these ISOCAM guaranteed time extra-galactic surveys (IGTES,
PI: Cesarsky) was to establish source counts in the mid-IR over two
orders of magnitude in flux (Elbaz et al. 1999). The largest field
surveyed during this program, already observed by the {\sl R\"ontgen
Satellite} (ROSAT, Hasinger \etal 1993), is known as Lockman Hole
since Lockman \etal (1986) pointed out the existence of this
exceptionally low HI absorption region. In the ISO data base, these
surveys have been complemented on the deep side by the surveys on the
 Hubble Deep Field North (Serjeant \etal 1997), the  Hubble Deep Field
South (Oliver \etal 2002), the  Canada-France Redshift Survey (Flores
\etal 1999), and by the lensed survey of Metcalfe \etal (2003), and on
the shallow side by the European Large Area Infrared Survey (ELAIS,
Oliver \etal 2000).

If we consider the typical  spectral energy distribution (SED) of a
star-forming galaxy (see e.g. Laurent \etal 2000), the presence of PAH
features makes it easily detectable by means of the LW3 band (centered
at 14.3~$\mu$m) of ISOCAM up to a redshift of 1.5.  In this redshift
range the ISOCAM data are, in terms of star formation, deeper than the
deepest radio surveys (see discussion in Elbaz \etal 2002).  Several
studies of optical spectra of local and high-z galaxies (Poggianti \&
Wu 2000, Poggianti, Bressan \& Franceschini 2001, Rigopoulou \etal
2000), found that more than 70\% of the energy emitted by young stars
and reprocessed in the far-IR leaves no traces in the optical spectra
(also after correcting for extinction). Even if a very refined
extinction correction using Hydrogen Balmer line ratios could provide
a correct estimate of the global star formation (Flores \etal 2004),
it would be impossible to compute that for any redshift since only a
few windows in the near-IR are accessible from the ground.

\begin{figure}
\centering
\includegraphics[width=0.5\textwidth]{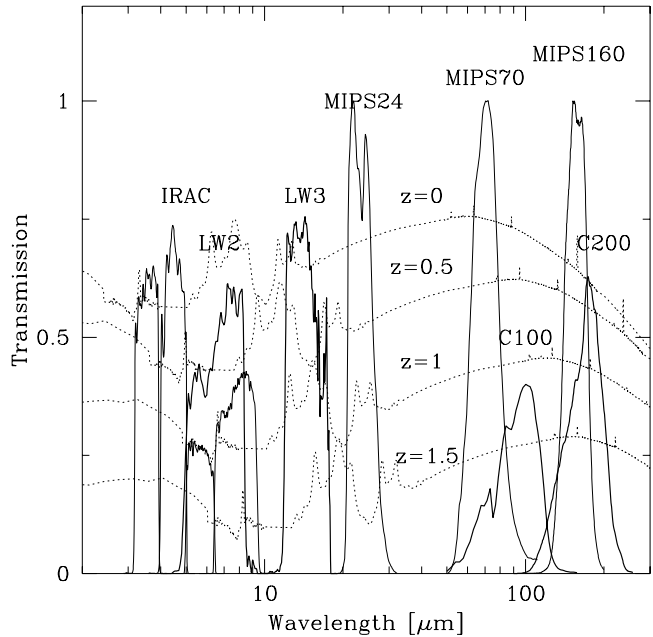}
\caption{Transmission curves of the Spitzer and ISO filters used to
survey the Lockman Hole on the M82 SED (dotted lines) put at the
redshift of z=0., 0.5, 1, and 1.5. The LW3 data remain unique also after the
Spitzer observations since the most important PAH features pass
through the LW3 filter in the redshift range $z=0-1.5$.}
\label{fig:filters}%
\end{figure}

As shown in Figure~\ref{fig:filters}, this band is essential to
distinguish between different types of starburst SEDs in the redshift
range $z=0-1.5$, since most of the prominent  mid-IR features (due to
Polycyclic Aromatic Hydrocarbons, PAH) lie in the wavelength range
covered by the LW3 band.  Even with the recently launched infrared
observatory Spitzer (Werner \etal 2004), the only possibility to
obtain images in this wavelength range is by using the peak-up camera of the
spectrograph with the blue filter (13.-18.5~$\mu$m) as an imager, but
its field of view is very small (80\asec$\times$56\asec).
The next space observatory able to observe in this wavelength region
will be the James Webb Space Telescope with its mid-IR
camera\footnote{http://ircamera.as.arizona.edu/MIRI/miriscience.pdf}.
Therefore, this set of data will be extremely useful for the
foreseeable future.

 In this paper we present the 14.3~$\mu$m image, catalogue and
identification of optical counterparts of the shallow ISOCAM survey in
the Lockman Hole.  The survey consists of 4 raster scans which
slightly overlap to cover 0.55 sq. degrees of sky for a total of 55
ksec observing time with the mid-infrared camera ISOCAM.  The
extension and depth of this survey is intermediate between the 
deep ISOCAM surveys (see e.g. Elbaz et al. 1999) and the shallow
and extended ELAIS survey (Oliver \etal 2000). This allows us to
obtain a sample of distant mid-infrared sources that is large enough
to study the deep universe without being biased by large scale
structures.  Moreover, the range of fluxes which is spanned by these
observations (0.5 - 4 mJy) covers the slope change in the 14.3~$\mu$m
counts discovered by Elbaz \etal (1999).  The sensitivity is limited
by the low redundancy of the survey, which is a necessary compromise
in order to survey large regions of sky.

Section 2 gives a summary of the ISOCAM observations and of the
optical follow-up of bright stars in the field for absolute
calibration purposes.  Section 3 describes the method used and
outlines the passages of the reduction and source extraction. The
absolute photometric calibration is also derived through a sample of
stars observed in optical, near-IR and 14.3~$\mu$m.  Section 4
discusses the accuracy of astrometry and photometry, as well as the
completeness of the survey, on the basis of simulations.  Section 5
describes the identification of optical counterparts of the 14~$\mu$m
sources and the catalogue is presented in Section 6.

In a companion paper (Rodighiero \etal 2004), we present the analysis
and catalogue of the 6.75~$\mu$m and 14.3~$\mu$m deep surveys in the
central region of the Lockman Hole and the 14.3~$\mu$m counts of the
two surveys combined.

\section{Observations}
\subsection{Infrared data}

\begin{table}
 \caption[]{Lockman Hole observation parameters.}
\label{tab:obsparam}
\begin{tabular}{l c}
   \hline
   \noalign{\smallskip}
   Parameter      &  Value  \\
   \noalign{\smallskip}
   \hline
   \noalign{\smallskip}
   Band effective wavelength             & 14.3~$\mu$m \\
   Band width                            &  6~$\mu$m \\
   Detector gain                         &  2 e$^-$/ADU \\
   Integration time                      &  5.04 s  \\
   Nr. of exposures per pointing         & 11   \\
   Nr. of stabilisation exposures        &115   \\
   Pixel field of view                   &  6\asec  \\
   Nr. of horizontal and vertical steps  &  24 $\times$ 8  \\
   Step sizes                            &  54\asec, 168\asec \\
   Nr. of raster maps                    &  4  \\
   Total area covered                    &  0.55 deg$^{2}$ \\
   \noalign{\smallskip}
   \hline
\end{tabular}
\end{table}

\begin{figure}
\centering
\includegraphics[width=0.5\textwidth]{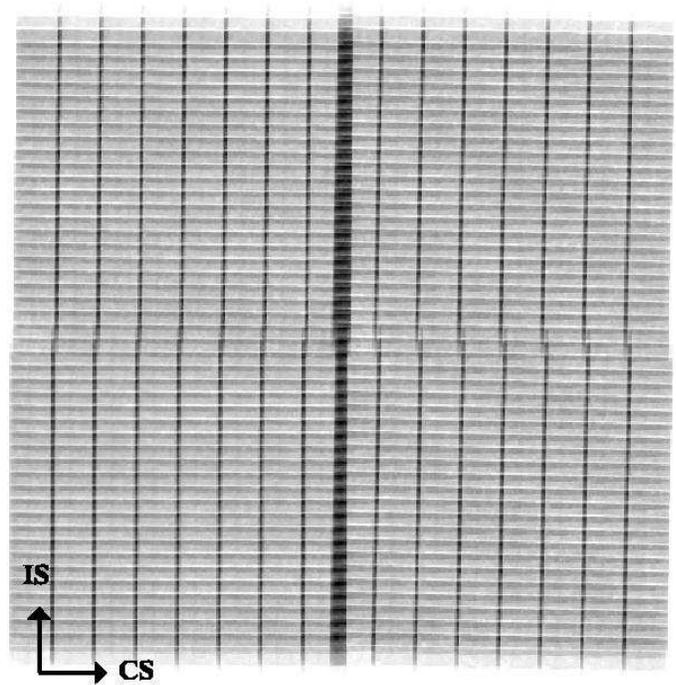}
\caption{Coverage map of the Lockman Hole shallow survey. Dark regions correspond
to the overlapping of raster scans ($\approx$ 350 seconds of exposure time), while
the typical coverage of grey regions is 200 seconds. Arrows indicate the in-scan (IS) and
cross-scan (CS) directions.}
\label{fig:coverage}%
\end{figure}

The data presented here correspond to the observations made during
orbits nr. 201 and 202 by ISOCAM in the direction of the Lockman Hole,
an area of the sky with low HI density:$N_H \sim 5 \times 10^{-9}$
(Lockman \etal 1986). ISOCAM spent a total exposure time of 55 ksec
observing a field centered in 10\h52\m03\s $+$57$^o$21\amin46\asec,
J2000.  Each of the four pointings which roughly cover a quarter of
the total field (observations nr. 20100901, 20101502, 20201003 and
20201104) correspond to a raster scan of $24\times 8$ sub-pointings,
with a mean number of 11 readouts of 5 seconds of exposure time. The
four rasters slightly overlap to cover the field without gaps and
obtain uniform coverage (see Figure~\ref{fig:coverage}). The mean
covering factor of the observation is 3.4, since the detector - an
array of 32 $\times$ 32 pixels - was displaced by 9 and 28 pixels along
and across the scan directions, respectively.
The choice of a 6'' pixel size is a compromise between angular resolution
and optimal signal-to-noise ratio (SNR) for the detection of faint sources.
The observation parameters are summarized in Table~1.

The same area has been also covered by other ISO observations: 90
~$\mu$m and 160~$\mu$m observations (Kawara \etal 1998, Rodighiero
\etal 2003) and deep 14.3~$\mu$m and 6.75~$\mu$m observations in the
$20' \times 20'$ central region (Rodighiero \etal, 2004).  Two Spitzer
programs (SWIRE, Lonsdale \etal 2003, and  guaranteed time
observations, P.I.: G. Rieke) have recently reobserved this region in
 all the Spitzer imaging bands.

\subsection{Ancillary data}

Several optical and near-IR images have been taken to support the
ISOCAM observations. In this paper we make use of a r' image taken
with the  2.5m Isaac Newton Telescope (INT) at La Palma which covers the
entire shallow survey. This image, along with other images in four
optical bands covering the center of this field, will be presented in
more detail in a forthcoming paper (Fadda, 2004).

Moreover, to check the absolute calibration of the source fluxes in
the survey, the central field has been observed in the U, B, g', r'
and i' bands with short exposures using the  INT Wide Field Camera
(WFC) during the nights of January 23, 2003 and March 27, 2003. Using
the 2MASS observations of the Lockman Hole  (Beichman \etal, 2003), we
gathered a sample of 21 stars emitting at 14.3~$\mu$m with J, H, Ks
and U, B, g', r', i' magnitudes to compute expected 14.3~$\mu$m
fluxes.  The bright stars, for which the 2MASS fluxes are more
uncertain, have been observed in the J, H and Ks bands with the 1.5~m
``Carlos Sanchez'' infrared telescope in Tenerife during the nights of
February 15 and 16, 2001.  These data have been reduced using
IRAF\footnote{IRAF, Image Reduction and Analysis Facility, is
distributed by the National Optical Astronomy Observatory, which is
operated by the Association of Universities for Research in Astronomy,
Inc., under a cooperative agreement with the National Science
Foundation} packages. In particular, the WFC data have been reduced
using the {\sl mscred} package developed for the analysis of mosaic
camera data and taking into account the information contained in the
web page of the INT Wide Field
Survey~\footnote{www.ast.cam.ac.uk/~wfcsur/index.php}.

\section{Data reduction}
\subsection{ The method}

\begin{figure}
\centering
\includegraphics[width=0.5\textwidth]{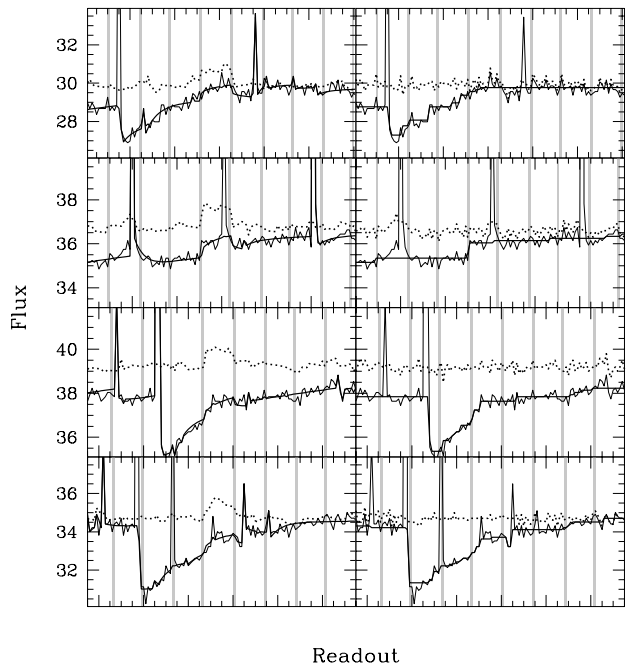}
\caption{Four examples of reduction of part of one pixel data with
the Lari \etal (2001, on the left) and the PRETI (on the right) methods.
Vertical lines separate the readouts for each pointing of the camera.
Raw data, model, and reconstructed signal are marked with thin, thick and
dotted lines, respectively. In these four cases the PRETI method fails
to detect the sources close to glitches or in negative parts of the signal.
}
\label{fig:lari_preti}%
\end{figure}

The impacts of cosmic rays on the ISOCAM detector have dramatic effects
on the pixel signals and make very difficult the detection of faint sources.
When a cosmic ray hits the detector a glitch appears in the signal which, 
depending on the energy of the hitting particle, could decay in a few readouts
or perturbate in a more serious way the signal up to several hundred readouts.

Detecting and correcting the pixel signals for the effects of these
cosmic rays is the main goal of the methods developed for the
extraction of faint sources from ISOCAM data.

The triple-beam technique by D\'esert \etal (1999) simply detects and
masks the regions affected by this transient behavior.  The Pattern
Recognition Technique for ISOCAM data (PRETI) by Starck \etal (1999)
decomposes the signal at different time scales and models the parts of
the signal below and over the median level. Then, it recognizes
patterns which are similar to sources and subtract the other parts
from the original signal.

The method by Lari \etal (2001) is the first which attempts to model
the signal using some physical hypothesis. It assumes that each pixel
has two charge reservoirs evolving independently with two different
time constants. Glitches due to cosmic ray impacts are treated like
discontinuities in the charges.  The signal observed $S$ is described
by the equations:
\begin{eqnarray}
S & = &  I + I_{dark} - \sum_{i=1, 2} \frac{dQ_i}{dt}  \\
\frac{dQ_i}{dt} &  =  & e_i (I + I_{dark}) - a_i Q_i^2  \qquad \qquad \textrm{ for } i=1, 2
\end{eqnarray}
where $I$ is the incident flux of photons, $I_{dark}$ the dark
current, $Q$ the accumulated charges. $e_i$ and $a_i$, parameters
describing the efficiency of the accumulation of charges and the time
constant, are estimated through a fit to the data.

This model works remarkably well (see examples in Lari \etal 2001)
allowing us to exploit in the best way the ISOCAM data. While the
triple-beam method simply does not consider the data affected by
transients losing the information contained in these readouts, the
non-parametric corrections done with PRETI can be sometimes dangerous.
PRETI in fact  does not consider the short glitches which are  smoothed before
starting the analysis. The multiscale transform is able to compute a
background (large scale) and detect positive and negative patterns in
the signal. The negative patterns are considered as negative tails
after cosmic ray impacts. The positive patterns are classified as
sources or positive tails after cosmic ray impacts.  This
classification is based on the temporal size of the pattern and on its
shape. Unfortunately, sometimes a positive tail is confused with a
source or, due to a bad evaluation of the local background, a positive
pattern can arise between two strong consecutive negative tails (see also
discussion in Fadda \etal 2000).  On
the other hand, a source in a negative tail will be not considered if
it lies under the local estimated background (see examples in
Figure~\ref{fig:lari_preti}).

These effects become important for faint sources, especially in the
case of data with low redundancy, such as the shallow survey in the
Lockman Hole or the ELAIS surveys (see e.g. Lari \etal 2001, Gruppioni
\etal 2002).

In conclusion, although the PRETI and the triple beam technique have
been successfully applied to data with high redundancy as, for
instance, the Hubble Deep Field North (Aussel \etal 1999, D\'esert
\etal 1999), our interactive analysis greatly improve the completeness
and reliability of faint sources in the case of low-redundancy data.

\begin{figure*}
\centering
\includegraphics[width=\textwidth]{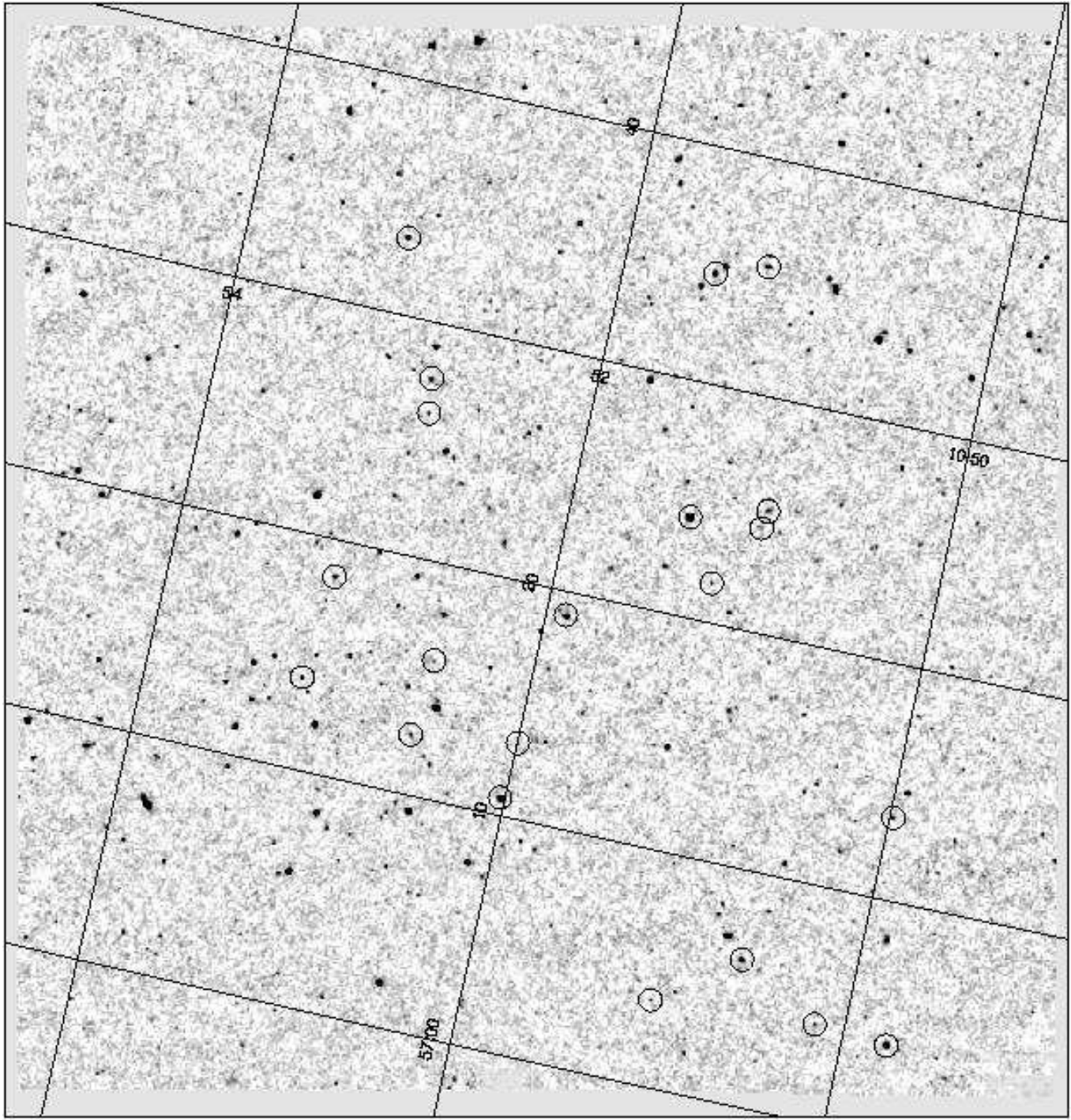}
\caption{SNR image of the total field observed by ISOCAM in the
direction of the Lockman Hole. The 21 stars used to check the
calibration flux factor are marked with circles.}
\label{fig:LH_map}%
\end{figure*}

\subsection{Reduction pipeline}

The CIA\footnote{CAM interactive analysis, developed by the ESA
Astrophysics Division and the ISOCAM Consortium. The ISOCAM Consortium
is led by the ISOCAM PI, C. Cesarsky. } package (Ott \etal 2001 ) was
used to build the raster structure from raw data and to subtract the
dark current from the data. The codes for data reduction with the
method by Lari \etal (2001) are written in the Interactive Data
Language (IDL) and the whole reduction and analysis is performed in
the CIA environment.

The first part of the reduction consists in the first estimation of 
background flux levels, identification of bright sources and glitches in each
pixel history. Hence, a fitting procedure is applied.
Subsequently, the signal is checked interactively where the fit failures
occurred or in noisy pixels. The small parts of the signal involved are refitted by
considering further glitches or sources not previously identified.
Although usually each pixel is treated individually, in cases of strong
cosmic ray hits the charges propagate in the surrounding pixels. To take 
into account also these cases, strong features from nearby pixels are 
considered in the fitting.

Once all the pixels are well fitted, the signal is flat-fielded and a
map is obtained with CIA routines which take into account the  geometric
distortion.  Positive and negative excesses in the map are
back-projected on pixel time histories and checked interactively,
eventually improving the fit of the pieces involved.

Hence, a further map is obtained and sources are extracted. We
consider all the sources at 4-$\sigma$ level which we back-projected
on the time pixel histories.  At this point, the last interactive
checking is done to recover all the faint parts of the source signal
if the source is real or to correct the fit if the source is false.
This last step allows us to improve the estimates of position and flux
of the sources since we better recover the wings of the  point spread
function (PSF).

\section{Mosaicing and source extraction}

Since the total observation is composed of four different rasters, 
the absolute astrometry of each raster map has to be corrected before coadding
the four maps in a unique image.
We used as astrometrical reference the r' image matching as many
objects as possible to correct for  shifts in the in-scan and
cross-scan directions. While shifts in the cross-scan direction are
around 2\asec, those along the in-scan direction are typically bigger ($\approx$ 7\asec).
After updating the astrometry information in the header of the
raster structure, we reproject all the images using the same scale and
orientation. Finally, the map is obtained summing the four maps by
weighting them with the exposure maps.

As discussed by Lari \etal (2001), it is better to consider the map
obtained without corrections for source transients ({\it unreconstructed image})
in the source extraction. In fact, the correction for the transient effects
does not work efficiently at low fluxes and it is better to correct for 
these effects with simulations. 

Since our extraction considers only the peak flux, we obtained better
results in the extraction using 6\asec~pixels sampled at a distance of 2\asec.
The image used for the extraction of point sources has a pixel
of 2\asec~size with the flux in each pixel corresponding to the flux recovered
by a 6\asec~pixel at the same center.

Another map of a 2\asec~pixel size has been obtained to measure the 
flux of extended sources. In this case, we used the flux reconstructed by the
model of Lari \etal (2001) which works well at high fluxes.

The  root-mean-square (RMS) image used for the detection is obtained
from the actual image and the exposure map.  The source detection is
done using the {\it find} routine of the IDL astronomical library (an
IDL implementation of a DAOPHOT routine). We check interactively all
the sources which have a peak in the SNR map higher than
4-$\sigma$.  Then, once corrections and new fits are done, we
reproject everything for the last time, extract the sources and retain
in the final catalogue only the sources at the 5-$\sigma$ level.
Errors in the source position and photometry depend on the source
SNR, as discussed further below.

\subsection{Source Photometry}

Measuring fluxes for ISOCAM faint sources is particularly challenging because
of the presence of strong memory effects after cosmic ray impacts. The presence
of these events in the pixel histories makes the photometry error extremely
variable across the image.
To overcome this effect, Lari \etal (2001) proposed to base the flux estimate
on the peak fluxes, at least for point sources, and to correct the measured
values for a scale factor deduced by simulating a source with a similar flux
at the location where the source was detected (the so called {\sl autosimulation}).
Since most of the sources are distant faint sources and the pixel field of view
is quite large (six arcseconds), the method is applied to almost every source
with a few exceptions. For extended sources aperture photometry is used.

Although the {\sl autosimulations} allow one to improve the
photometric accuracy, the fact that the actual position of the source
is poorly known typically leads to underestimates of the actual flux. To
take into account this effect, we run extensive simulations which also
allow us to estimate the errors and the completeness of the survey (see further below).

Finally, to transform instrumental units to physical units, one can
refer to the factors computed by Blommaert \etal (2000) with
calibration stars during the ISO mission.  We have checked this
calibration with a few stars in our field for which we collected
near-IR and optical magnitudes.

\subsection{Absolute calibration}

Blommaert \etal (2000) calibrated the ISOCAM detectors using a few
stars, observed several times during the ISO mission, for which
detailed SEDs were available. In the case of the LW3 band, seven stars
have been used.  Although the computed sensitivity factor is in good
agreement with pre-launch values, a large scatter is present in the
data. This could depend on the reduction method used (especially for
transient correction) or on possible variations of the detector
 responsivity during the mission.

Since in our field there are many bright stars emitting in the LW3 band, 
we decided to independently compute an absolute calibration factor by
observing some stars in near-IR and optical bands and fitting their SEDs
with Kurucz (1993) stellar models. Although the spectral type of these stars is unknown, 
this method has the advantage of being absolutely coherent (stars and
extra-galactic sources are reduced in the same manner) and independent of the
variations of the detector responsivity during the mission (stars and
extra-galactic sources were observed at the same time).

For this reason, we observed 21 stars in the J, H and Ks near-IR bands
and in the U, B, g', r' and i' optical bands.  We fitted the SEDs of
these stars using a grid of 253 Kurucz (1993) models to deduce the
expected flux at 14.3~$\mu$m.  Table~\ref{tab:calstars} summarizes the
photometric data and the expected LW3 fluxes (LW3$_{exp}$). All the magnitudes are
Vega-like and an asterisk indicates that the star is slightly
saturated in the band. Optical magnitudes have been derived using the
{\it mag\_auto} magnitude of SExtractor (version 2.3; Bertin \&
Arnouts 1996).  The estimates take into account the saturated pixels.
Lower weights have been assigned to magnitudes of slightly saturated stars 
in the least-square-mean fits to the Kurucz  (1993) models.

\input{LHSStab2.tex}

\begin{figure}
\centering
\includegraphics[width=0.5\textwidth]{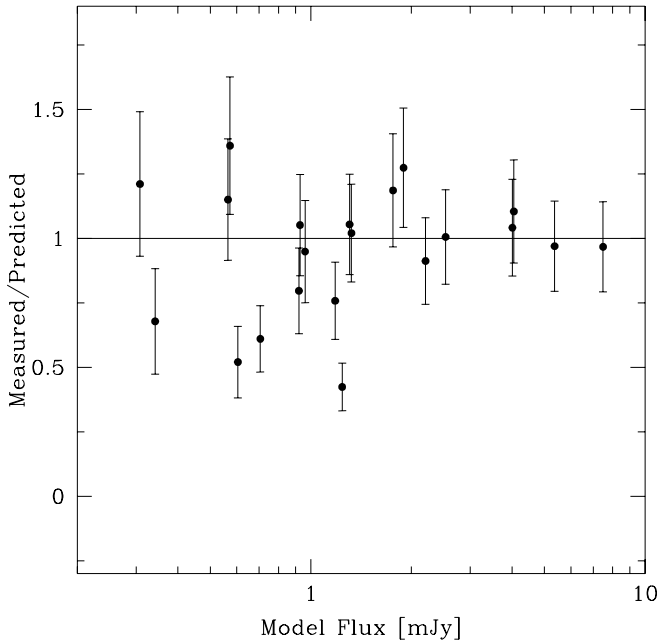}
\caption{The ratio between measured and predicted  fluxes for 21 stars detected
at 14.3~$\mu$m in the Lockman Hole shallow survey. Errorbars take into account
the error on the photometry and the error in the predicted fluxes. The median
of the points is 1.0.}
\label{fig:stars}%
\end{figure}

As shown in Figure~\ref{fig:stars}, we found that the median ratio
between predicted and measured fluxes (assuming the calibration factor
by Blommaert \etal, 2000) is 1.0. Since most of the points lie inside
a $\pm 25\%$ band, which correspond to our photometric errors, we
conclude that this calibration factor is in agreement with our study
and can be safely assumed also in case of faint fluxes.  This is a
very important point since these catalogues are used to compute deep
14.3~$\mu$m extra-galactic counts.
The counts of Gruppioni \etal (2002) assume, for instance, a slightly
different calibration based on a relationship between near-IR magnitudes
and IRAS fluxes.

\begin{figure}
\centering
\includegraphics[width=0.5\textwidth]{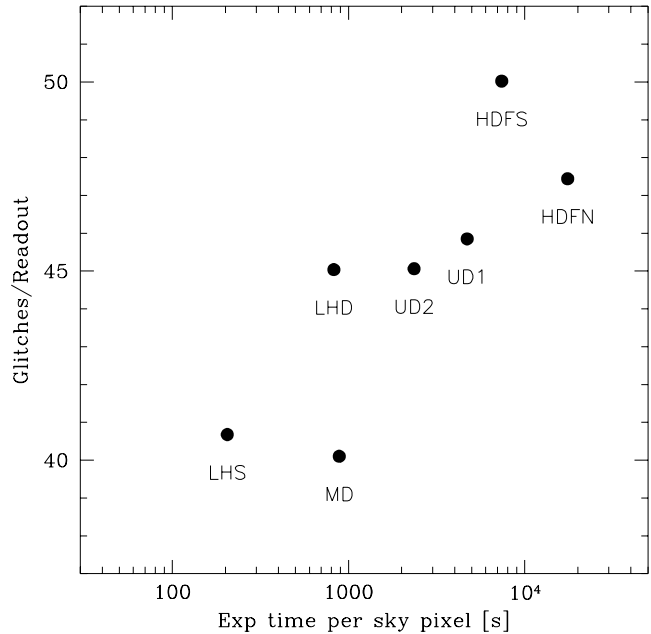}
\caption{ Number of glitches per frame versus exposure time per sky
pixel for the ISOCAM deep surveys (LHD and LHS, Lockman Hole Deep and
Shallow, HDFN and HDFS, Hubble Deep Field North and South, MD, Marano
Deep, UD1 and UD2, Ultra-deep fields in the Marano field).  The number
of glitches in the signal due to cosmic ray hits are computed for
glitches which are 5 times bigger than the rms of the signal for each
pixel. The surveys with the longest scans (LHS and MD) have the lowest
number of glitches.}
\label{fig:surveys}%
\end{figure}

\section{Survey Performance}

In spite of its low redundancy (on average three images for each sky
point), the Lockman Hole shallow survey is one of the highest quality
ISOCAM extra-galactic surveys. As visible in Figure~\ref{fig:coverage},
the coverage is highly homogeneous and without any gap. Moreover, each
raster scan has the highest number of readouts among the ISOCAM
surveys. Because of the presence of a long term transient in the
ISOCAM data, this allow the detector to stabilize and have a more
predictable behavior on a large part of the scan.  Finally,
observations have been performed during a period of low cosmic ray
flux with respect to other deep surveys. In Figure~\ref{fig:surveys},
we compare several deep ISOCAM surveys which have been performed with
the same gain, exposure time per readout and pixel field of view of
the Lockman Hole shallow survey.  We evaluated the rate of cosmic rays
by counting the number of glitches in the pixel signals beyond 5 times
the rms of the signal.  Since the transients caused by cosmic ray hits
are the most difficult thing to correct in the signals, this quantity
gives a direct measure of the quality of the data.

Therefore, in spite of the low depth of these data with respect to
other ISOCAM surveys, the low rate of cosmic rays and the long scans
used in the observations make the Lockman Hole Shallow Survey one
of the best extra-galactic surveys performed by ISOCAM.

The astrometric and photometric accuracy, as well as the completeness
limits of the survey have been estimated through a set of simulations
at different flux levels. We describe in the following how the
simulations have been performed and analyzed. Moreover, since the
central region has been observed twice at different depths, we discuss
also the photometric accuracy and source extraction reliability
comparing our catalogue to that of the deep survey (Rodighiero \etal
2004).

\subsection{Simulations and Completeness}

Although we autosimulate the extracted sources to recover the flux 
from the wings of their PSFs, a small fraction of flux remains undetected
since the center of the source is not precisely known.
To estimate statistically this correction, as well as to study the errors
in astrometry and photometry and the completeness limit of the survey, 
simulations are needed.
We decided to perform a set of simulations at several fluxes
(0.35, 0.5, 0.7, 1, 1.4, and 2 mJy) which span the entire flux range of
the survey.

Each simulation has been done introducing, in one of the rasters, a
set of 50 artificial sources in regions observed at least 100 seconds
to exclude the noisy borders of the images. The sources have been put in 
positions with SNR less than 2 in order to avoid 
overlapping real sources. This synthetic image, created using the LW3
PSF, has been backprojected to the data cube adding the transient behavior
according to the model of  Lari \etal (2001) and taking into account the
camera distortions and flat-fielding.
Finally, the synthetic data cube added to the original raw data cube
has been reanalyzed in the same way as the original cube. 
To speed up the process, the interactive part of the reduction has
been applied only to the parts of the cube close to the added artificial
sources.

\begin{figure}
\centering
\includegraphics[width=0.5\textwidth]{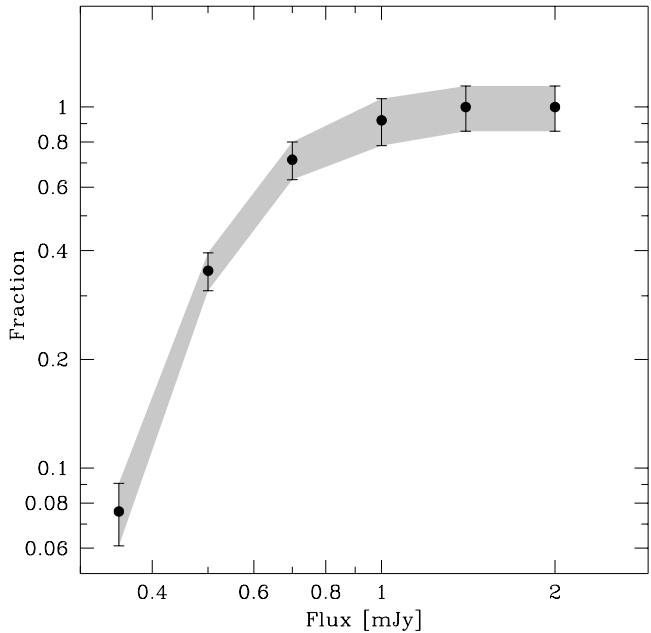}
\caption{Fraction of simulated sources detected as a function of the input flux.
Bars correspond to the 1-$\sigma$ Poissonian errors. The survey is 50\% complete
at 0.6 mJy
}
\label{fig:completeness}%
\end{figure}

To have enough data for reliable statistics at each flux, we performed seven simulations
at 0.35 mJy, four at 0.5 mJy, three at 0.7 mJy and one for the other
fluxes.  The number of sources detected at each flux automatically
gives us the estimate of the completeness (see
Figure~\ref{fig:completeness}).  The survey is therefore close to
100\% complete for fluxes brighter than 1 mJy.  The 20\%, 50\%, and
80\% completeness limits are 0.45, 0.6, and 0.8 mJy, respectively.

\begin{figure}
\centering \includegraphics[width=0.5\textwidth]{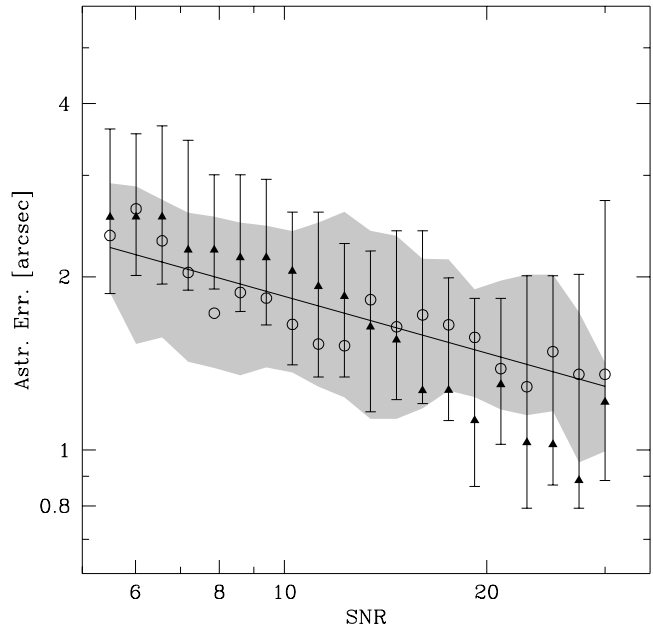}
\caption{Astrometric accuracy versus SNR for
simulated sources (empty circles) and real sources (filled triangles)
compared to their optical counterparts. The points correspond to
1$\sigma$, i.e. the distance inside which one finds 68\% of the counterparts.
The lower and upper limits (errorbars for the real sources and shaded area
for the simulated ones) correspond to the distances inside which 
one finds 50\% and 80\% of the points, respectively. The distances
of the simulated points have been quadratically added to the pointing
accuracy.}
\label{fig:err_vs_snr}%
\end{figure}

\begin{figure}
\centering
\includegraphics[width=0.5\textwidth]{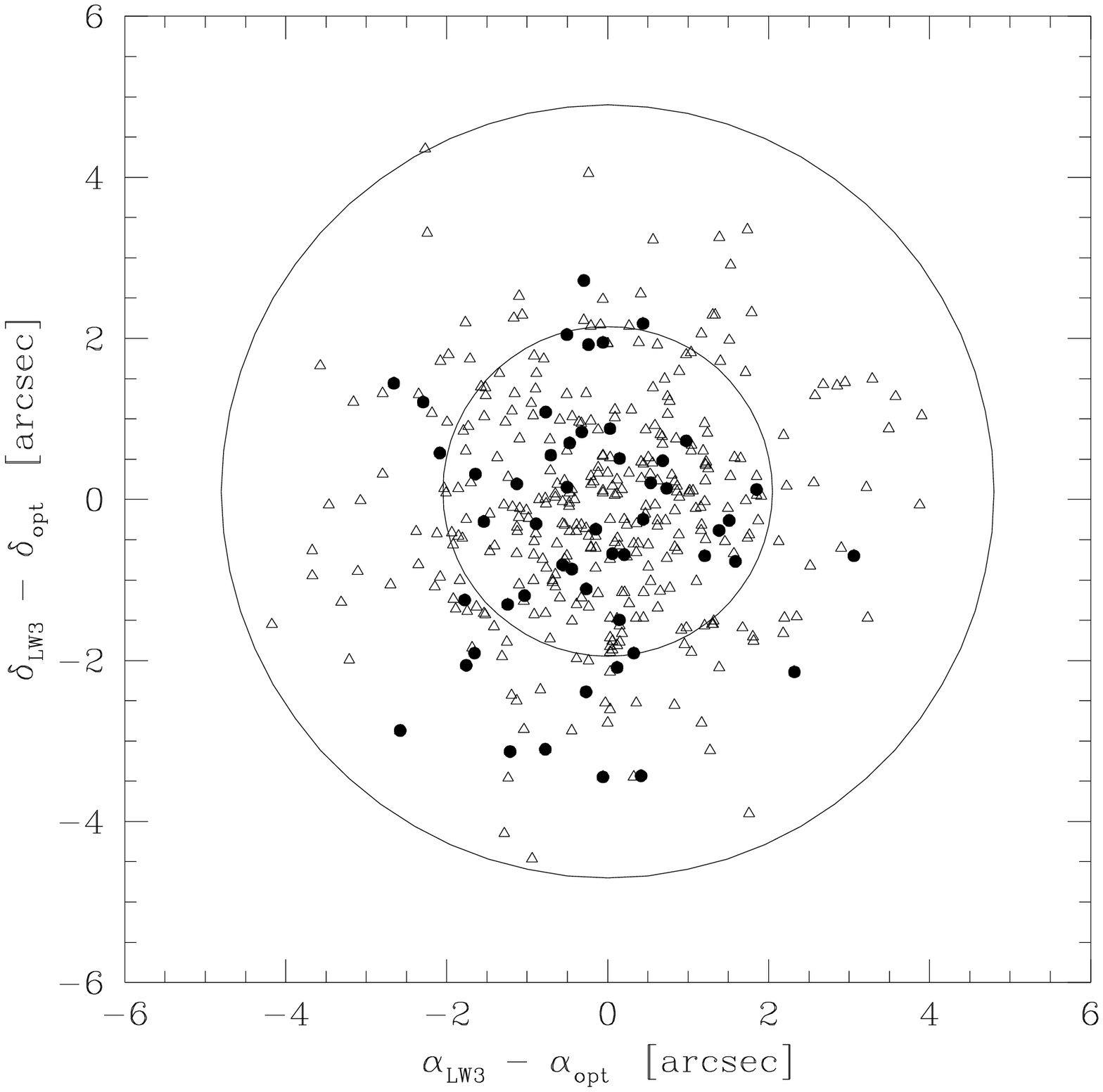}
\caption{Difference in position between the infrared sources and 
their optical counterparts. The diameter of the external circle 
corresponds to the LW3 beam (4.7\asec), while the internal
circle contains 68\% of the points (2\asec). The stars
are indicated with filled dots.
}
\label{fig:offsets}%
\end{figure}

\subsection{Astrometric Accuracy}

To evaluate the astrometric accuracy we used the simulations and
the cross-correlation with the optical image.
While the comparison with the optical image gives a direct estimate
of the astrometric accuracy, one has to add to the error estimated
from the simulations the pointing error.
In fact, although we improved the absolute astrometry of the LW3 images
adding the offsets between the brightest LW3 sources and their optical
counterparts, the dispersion of the estimated offsets is still significant (around 0.6\asec).

In Figure~\ref{fig:err_vs_snr} we report the astrometric error as a
function of the SNR of the detection for the
simulated and observed sources. To show the uncertainty in the
measurement of the astrometric error, we plotted the radii containing
at each value of SNR 50\%, 68\% and 80\% of the counterparts (true and
optical positions for the simulated and observed sources, 
respectively).  The agreement between simulations and
cross-correlation of optical and real sources is generally good, 
except for high SNR values where the statistics of the real sources
are poor.  The line, obtained through a least square fit of the
simulated points, traces the 1-$\sigma$ error reported in our
catalogue. The typical error is lower than 2.5\asec~and, for well
detected sources with SNR $> 10$, is less than 2\asec.

The $\alpha$ and $\delta$ offsets between the observed sources and their optical
counterparts are shown in Figure~\ref{fig:offsets}. Within a circle of 2 arcsec
we have 68\% of the optical counterparts, i.e. 1-$\sigma$ in the case of a Gaussian
distribution. We have considered in this case, only optical counterparts inside
the beam of the LW3 image (4.7\asec).

\begin{figure}
\centering
\includegraphics[width=0.5\textwidth]{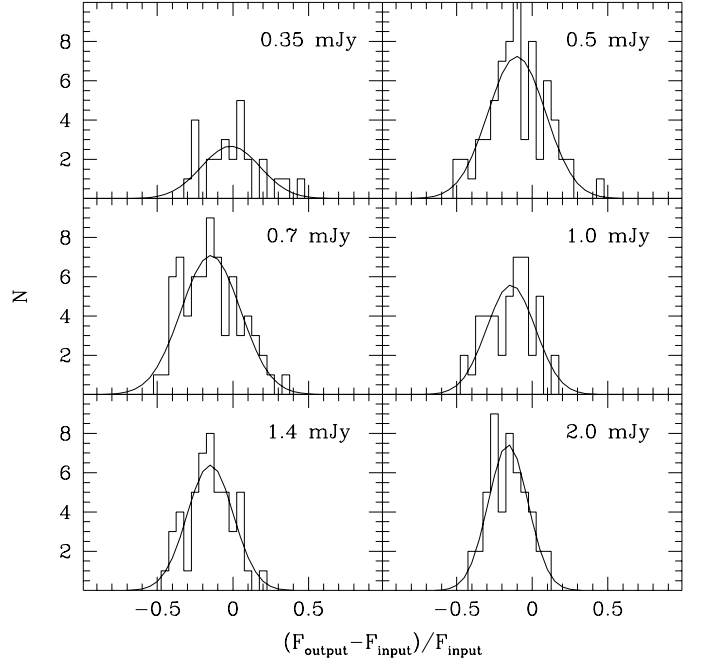}
\caption{Distribution of fluxes of detected sources with respect to the input flux of the simulated sources.
Gaussian curves have parameters computed by means of a biweight estimator.
}
\label{fig:simphot}%
\end{figure}
\begin{figure}
\centering
\includegraphics[width=0.5\textwidth]{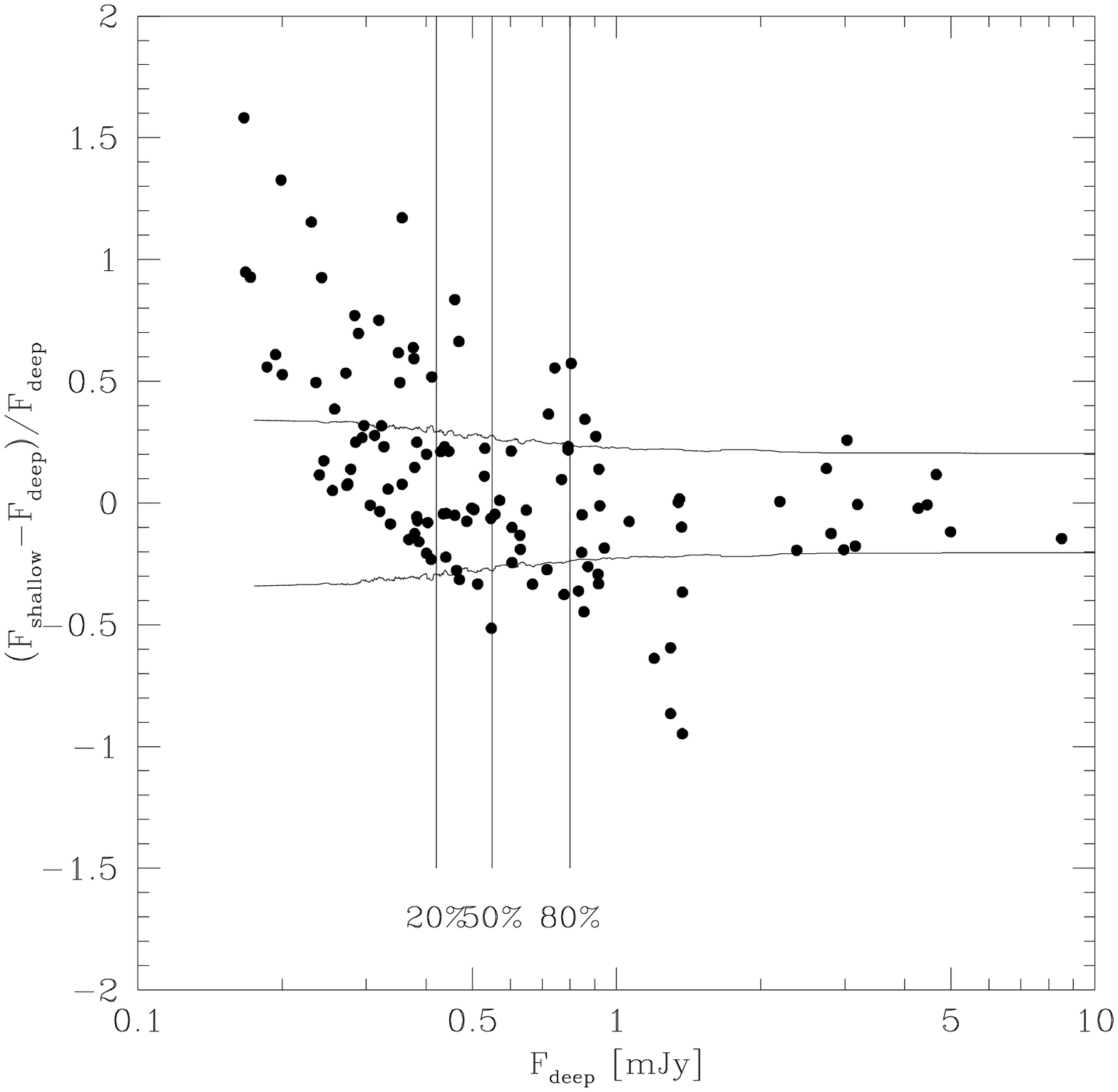}
\caption{Comparison of flux estimates for sources detected in both the shallow
and deep Lockman Hole surveys. The caustics represent the 1$\sigma$ photometric errors.
Below the 20\% completeness limit, we detect an increasing number of faint sources enhanced by
positive oscillations of the noise.
}
\label{fig:photometry}%
\end{figure}

\subsection{Photometric Accuracy}

The photometric accuracy of the survey has been studied through our
set of simulations. Moreover, since the central part of the field has
been reobserved deeply a second time (Rodighiero \etal 2004), we 
also compared the photometry of the sources common to the two surveys.

Figure~\ref{fig:simphot} summarizes the results of our set of simulations.
The location and dispersion of the distributions have been evaluated by
means of the biweight estimator (Beers, Flynn, \& Gebhardt 1990).
Except for the case of very low fluxes (0.35 mJy), the flux measured is
always lower than the true flux. The median ratio between output and input fluxes
is stable for fluxes greater than 0.5 mJy converging to the value of 0.84 at 2 mJy.
We assume this value to correct the bias in the flux measurement of real sources.

Going towards low fluxes, the detection is more affected by the structure 
of the noise. Sources on the top of positive oscillations of the noise are
enhanced allowing their detection, but also affecting their flux measurement.
We remind in fact that our estimates of the flux are based on the peak value
which is highly biased in these cases.

This effect is clearly visible in Figure~\ref{fig:photometry} where we compare
the fluxes of the same sources detected in the shallow and deep surveys.
In the case of faint sources, the flux estimate based on the shallow survey
is overestimated with respect to that based on the deeper observation.

Therefore, the fluxes of sources fainter than the 20\% completeness
limit are typically overestimated and it is very dangerous to consider sources
below this limit for counts purposes. 
When used for computing SEDs, these values have to taken as upper limits of the
flux.

As discussed in Lari \etal (2001), our photometric errors come from the 
autosimulation process and the noise in the sky map:
\begin{equation}
(\Delta S/S)^2 = \Delta (s_{out}/s_{in})_{autosim}^2 + SNR^{-2}.
\end{equation}
The second term is very important at low fluxes and negligible for
high SNR sources.  So, we can derive the first term from our
simulations at high fluxes. Considering only sources detected with
$SNR > 25$, the first term is 0.14.  At our lower limit of $SNR=5$, 
the relative error is 25\% and converge to 15\% at $SNR>20$.

\subsection{Reliability}

Because of the low redundancy and the difficulty in properly correcting
the transients caused by energetic cosmic rays, we expect a certain
fraction of the sources in our catalogue to be spurious detections.  A
small percentage might be also due to detections of transient events
(asteroids or variable objects).

\begin{figure}
\centering
\includegraphics[width=0.5\textwidth]{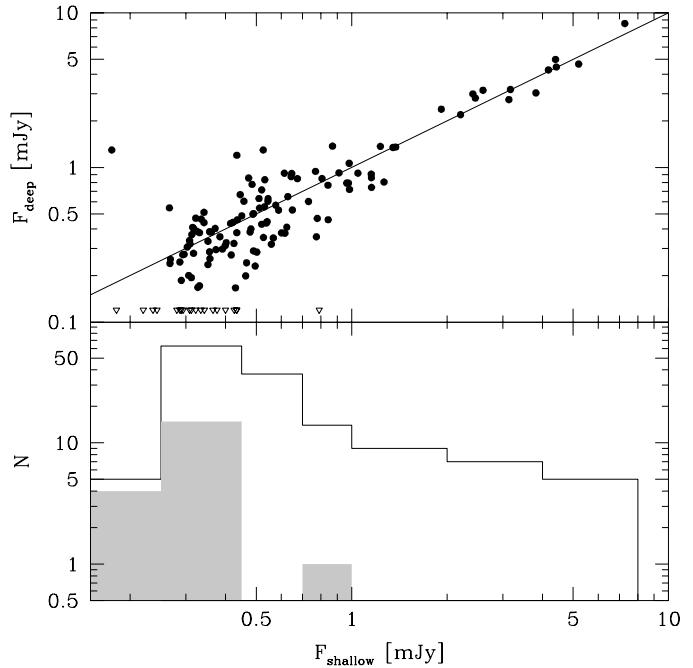}
\caption{Comparison between sources detected in the central region of the
field by the shallow and deep surveys. Top panel: black dots are sources
detected in the two surveys while empty triangles only in the shallow one.
Bottom panel: total number of sources in the shallow survey (solid line)
and shallow detection not present in the deep survey (shaded histogram).
Sources brighter than 0.45 mJy have a high degree of reliability.
}
\label{fig:reliability}%
\end{figure}

Since  a quarter of the area has been reobserved at the same
wavelength in a deeper survey, we can match the sources in our catalogue
with those of the deep survey (Rodighiero \etal 2004) to evaluate the
false detection rate.

We stress here that this analysis gives an upper limit to the number
of spurious detections, since some of the faint sources detected in
the shallow survey can be missed by the deep survey. In fact, at faint
fluxes, also the deep survey is not 100\% complete and a certain
percentage of sources is missed mainly because of the effect of
uncorrected cosmic rays.

141 sources of our catalogue fall in the region of the deep survey.
Using a matching radius of 9\asec (twice the  full-width half maximum
(FWHM) of the 14.3~$\mu$m PSF), we find 121 matches with the deep
survey catalogue, i.e. 86\% of the detections have a counterpart in
the deep survey.  However, as shown in Figure~\ref{fig:reliability},
the probable spurious detections have very low fluxes (lower than 0.45
mJy, i.e. the 20\% completeness limit).

We do not consider in our final catalog sources fainter than 0.25~mJy
which are almost certainly false detections. 

Between 0.25~mJy and 0.45~mJy, the percentage of false detections is 24\%
and it decreases to a very low level at brighter fluxes.
For this reason we split our catalogue into two lists.
The first one presents the 260 sources with flux greater than 0.45~mJy
which are reliable and have unbiased photometry.

Because of the great scientific interest of faint infrared sources, we
also provide a supplementary list of 197 sources with estimated flux
lower than 0.45~mJy with the caveats that the percentage of false
detections is high (24\%) and that some of sources have overestimated
fluxes. The 16 sources which are undetected in the deep surveys are
commented with a question mark in the catalogue. 10 of them have no
optical counterpart or a probability of random association bigger than
10\%.

\section{Optical counterparts}

We searched for optical counterparts on a r' image which covers the
entire ISOCAM survey. In this paragraph, we describe the optical image
and the method used for the identification.

\begin{figure}
\centering
\includegraphics[width=0.5\textwidth]{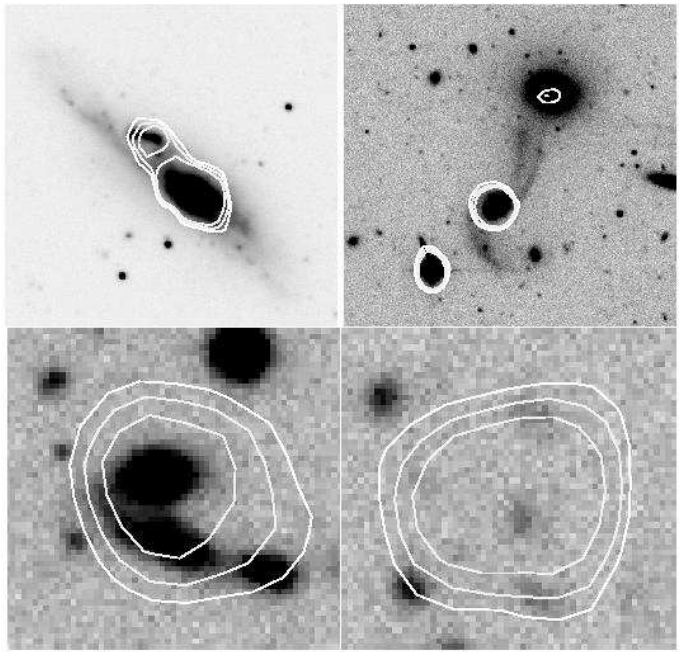}
\caption{Four optical counterparts of 14.3~$\mu$m sources. SNR levels of 5, 7, and 10 have been overlapped to the r' image.
The two images on the top have a field of 2\amin$\times$2\amin, 
the two on the bottom of 20\asec$\times$20\asec. In the top left
corner the only IRAS source in the field, NGC 3440, which is
resolved in  two components. At the top right and bottom left, two examples of interacting galaxies.
At the bottom right, faint optical counterpart of a strong mid-IR source.
}
\label{fig:counterparts}%
\end{figure}

\subsection{The optical image}

The optical image was taken in the Sloan r' filter band at the Isaac
Newton telescope (INT) in La Palma, Spain in two nights (2001, Dec. 12 and
2002, Jan. 20).  The field was covered with four pointings with five dithers
at each position.  Moreover, to cover a
small part of the ISOCAM field which was out of the observed field, we
used also an archival image taken as part of the Wide Field Survey
with the INT (McMahon \etal, 2001).

The images have been reduced using IRAF packages (in
particular {\it mscred}) and taking into account the non-linearity of the CCD
response and the radial distortion term described in the INT Wide Field Survey
web page~\footnote{www.ast.cam.ac.uk/~wfcsur/index.php}.
The astrometry has been added taking as a reference the GSC-II catalogue\footnote{The Guide Star
Catalogue-II is a joint project of the Space Telescope Science
Institute and the Osservatorio Astronomico di Torino. Space Telescope
Science Institute is operated by the Association of Universities for
Research in Astronomy, for the National Aeronautics and Space
Administration under contract NAS5-26555.  The participation of the
Osservatorio Astronomico di Torino is supported by the Italian Council
for Research in Astronomy.  Additional support is provided by European
Southern Observatory, Space Telescope European Coordinating Facility, 
the International GEMINI project and the European Space Agency
Astrophysics Division.}.

To improve the background matching and obtain a smooth constant background, 
the bright Tycho stars in the field have been subtracted from each image.

The photometric zero-point has been evaluated using the standard stars
in the night with the best transparency and every image has been scaled
to an image taken during this night. Finally, the images have been
coadded to obtain an  image covering the entire LW3 field.

 The spatial resolution is variable across the image
(0.9\asec $<$ FWHM $<$  1.3\asec) due to different seeing conditions.
Since the RMS of the images is slightly different for the different
nights, the limiting magnitude depth varies from r'=25 to r'=25.5
(Vega), as measured at 5-$\sigma$ inside an aperture of
1.35$\times$FWHM of a stellar PSF (optimal aperture to include most of
the flux and least of the background in the case of a Gaussian PSF).

A catalogue of sources has been produced using SExtractor (version
2.3; Bertin \& Arnouts 1996) using a 3$\times$FWHM aperture magnitude
and the {\sl auto\_mag} for extended sources. In the catalogue we
considered only sources with SNR greater than 3
within an aperture of 1.35$\times$FWHM.  Since bright stars are
saturated in this image, we will report in the catalogue, when this is
possible, the magnitudes from the shallow exposures taken for
calibration purposes. Position of saturated stars are taken from the
Tycho2 catalogue.

 SExtractor computes also a stellar index which can be safely used for
magnitudes brighter than r'=23 to separate star-like objects from
galaxies.  We have used this index to identify stars in our catalogue
(values greater than 0.85), unless in case of saturated stars which
are easily classifiable by direct inspection of the image.

More details about the data reduction and source extraction of this
optical image, as well as of a set of images in other four optical
bands observed in the center of the same field, will be given in a
forthcoming paper (Fadda, 2004).

\subsection{Identification of the counterparts}

To search for optical counterparts of the mid-IR sources, we have
considered a maximum distance of 4.7\asec, i.e. the FWHM of the 14.3~$\mu$m
PSF. As shown in Figure~\ref{fig:offsets}, most of the sources lie inside
a 2\asec radius circle which agree very well with the typical astrometric
error of the LW3 sources. 

In general, mid-IR sources have clear optical counterparts. In a few
cases, they correspond to a pair of interacting galaxies or there are
several possible counterparts (see Figure~\ref{fig:counterparts} for some
examples). In the entire field there is only one source which has been
detected by IRAS (the galaxy NGC 3440) and that has been resolved by ISO in two components.
This is also the only extended source in our survey.

For each source, we have computed the probability of random association
of the mid-IR source with its optical counterpart. Assuming a Poissonian
distribution of the optical sources, 
\begin{equation}
P = 1 - e^{-n(r') \pi d^2}
\label{equ:prob}
\end{equation}
gives the probability of random association within a distance $d$ with
optical sources brighter than $r'$, where $n(r')$ is the expected number density
of optical sources brighter than $r'$ (the magnitude of the possible counterpart).
To evaluate $n(r')$ we used the counts of sources in our optical image.

\begin{figure}
\centering
\includegraphics[width=0.5\textwidth]{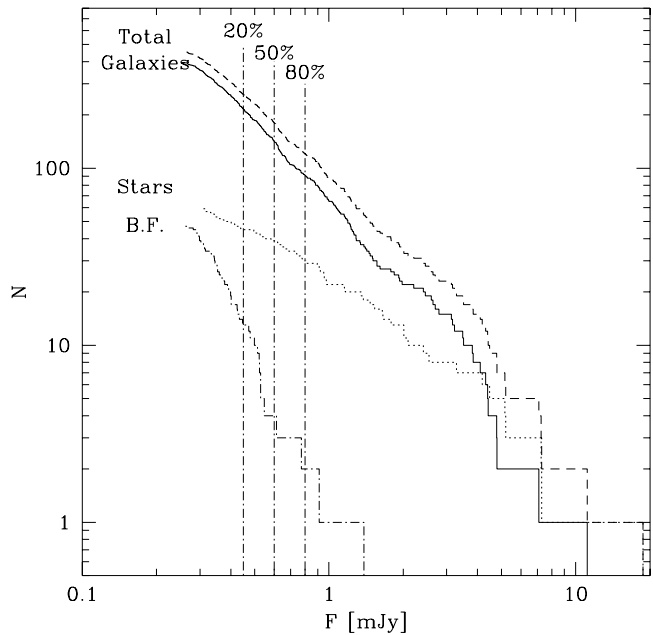}
\caption{Cumulative counts for sources, galaxies, stars and blank fields
are marked with dashed, solid, dotted and dash-dotted lines, respectively.
For fluxes greater than 0.6 mJy (50\% completeness limit), 21\% of the 
sources are stars and less than 2\% are blank fields in our r' image.
}
\label{fig:cumulative}%
\end{figure}

Figure~\ref{fig:cumulative} describes the cumulative counts of the
sources, galaxies, stars and blank fields, i.e. sources without
optical counterparts in our image.  Stars are 21\% of the sources for
fluxes greater than the 50\% completeness limit (0.6 mJy). The number
of blank fields is very limited for sources brighter than the 20\%
completeness limit (less than 5\%) and is less than 11\% for the all
sources. Since under the 20\% completeness limit the number of false
detections increases very rapidly, many of the blank fields correspond
probably to false detections.

\section{Data Products}

Images in fits format and catalogues in ASCII format are made
available to the astronomical community through the
world-wide-web~\footnote{via anonymous ftp to cdsarc.u-strasbg.fr
(130.79.128.5) or via http://cdsweb.u-strasbg.fr/cgi-bin/qcat?J/A+A/,
http://spider.ipac.caltech.edu/staff/fadda/lockman.html and
http://irsa.ipac.caltech.edu/data/SPITZER/SWIRE}
or directly on request from the authors.

\subsection{Images}

The images which are made publicly available have a size of 7.6 Mbyte
and consist of:

{\bf Flux map}: an image with pixels of 2\asec~and units of mJy/pixel.
This image corresponds to the reconstructed image, i.e. the image with
transients of the sources corrected by the model. We stress that, in 
case of faint fluxes, the reconstruction does not work properly and that
we computed the fluxes using the uncorrected image and correcting the
fluxes with autosimulations;

{\bf Coverage map}: an image with pixels of 2\asec~and units of number
of readouts. Since every readout has an integration time of 5.04 s, the
exposure map can be obtained simply multiplying this map by the integration time
per readout;

{\bf SNR map}: this map has a resolution of 2\asec and has been used
for the extraction of source positions. Every pixel contains the SNR
of an equivalent pixel of 6\asec~(the natural beam of the
observations) with the same center.  Thanks to this  kind of
resampling, we have a better measurement of positions and fluxes.

\subsection{Catalogue}

The catalogue is split in two parts: the first one containing 260 highly
reliable detections with estimated fluxes greater than 0.45~mJy and 
a second one with the 197 sources fainter than this limit for which the 
rate of false detections is around 24\%.
The two Tables (\ref{tab:catalog1} and \ref{tab:catalog2}) list:

\begin{itemize}
\item Column 1: the full designation of the source  recommended by the
International Astronomical Union (IAU). The prefix is composed by the
name of the satellite (ISO) and that of the survey (Lockman Hole
Shallow Survey, LHSS);
\item Columns 2-3: right ascension and declination (J2000);
\item Column 4: estimated astrometric error in arcsec;
\item Column 5: 14.3~$\mu$m flux and  respective error in mJy;
\item Column 6: redundancy (number of frames) of the detection;
\item Column 7: peak SNR of the detection;
\item Column 8: r' magnitude of the optical counterpart in the Vega system;
\item Column 9: SNR of the optical detection computed within an aperture of 1.35 $\times$ FWHM;
\item Columns 10-11: right ascension and declination offsets of the optical counterpart in arcsec;
\item Column 12: probability of random association, computed according to the equation (\ref{equ:prob});
\item Column 13: notes about the object: star, if the object corresponds to 
a star (in the case the star is saturated on our image, we report the Tycho2 position);
pair, if the IR emission comes from a pair of interacting galaxies; bridge, in one case
the IR emission seems to come out of a faint bridge between two interacting galaxies;
question mark, the source has not been detected in the deep survey (Rodighiero \etal 2004).
\end{itemize}

\section{Summary}

The Lockman Hole Shallow Survey, the shallowest and most extended
among the IGTES surveys, has been reduced with the technique of Lari
\etal (2001).  These ISOCAM 14.3~$\mu$m observations cover a region of
0.55 square arcminutes in the direction of the Lockman Hole.  457
sources are detected above the 5-$\sigma$ threshold with fluxes in the
interval 0.25-19. mJy.

 Completeness and photometry accuracy of the catalogue have been
assessed through a series of simulations at different flux levels.
The survey is 80\% complete at 0.8 mJy and 50\% complete at 0.6 mJy.
The positional accuracy, estimated with simulations and
cross-correlation of infrared and optical sources, is around
1.5\asec~for objects with SNR greater than 20 and
around 2\asec~for objects with SNR close to 5.  We have checked also
the absolute calibration using a set of 21 stars in the field observed
in near-IR and optical bands. Our analysis confirms the factor
computed by Blommaert \etal (2000).  From the comparison with the deep
survey in the same region (Rodighiero \etal 2004) we conclude that
most of the spurious detections have fluxes lower than 0.45 mJy (also
the 20\% completeness limit of the survey).

 Within the limiting depth of our optical image of the field (r'$=25$),
we find 95\% of counterparts for sources brighter than 0.45 mJy (the
20\% completeness limit) and 89\% in total. Stars make up 21\% of the
sources for fluxes brighter than 0.6 mJy (the 50\% completeness limit)
and 13\% in total.

In a companion paper (Rodighiero \etal 2004), we present the analysis
of the deep survey in the central region of the Lockman Hole at
14.3~$\mu$m and 6.75~$\mu$m and the combined 14.3~$\mu$m counts.  Forthcoming
papers will present the imaging and spectroscopic follow-up
observations of these fields and a cross-correlation between infrared
and radio sources.

\begin{acknowledgements}
Part of this work was supported by the ``Spanish Ministerio de Ciencia
y Tecnologia'' (grant nr. PB1998-0409-C02-01) and by the EC Network
"POE" (grant nr. HPRN-CT-2000-00138).  D.F. thanks Lisa
Storrie-Lombardi for her comments and careful reading of the
manuscript. He is also grateful to Yves Grosdidier who introduced him
to the TCS telescope.

\end{acknowledgements}

\clearpage

\input{LHSStab3.tex}

\clearpage

\input{LHSStab4.tex}

\end{document}

%% file: LHSStab2.tex
%

\begin{table*}
\caption[]{Calibration stars.}
\label{tab:calstars}
\begin{tabular}{cc ccccc ccc cc}
   \hline
   \noalign{\smallskip}
  RA   &    DEC  & u & B & g' & r' & i' & J & H & Ks & LW3   & LW3$_{exp}$ \\
(J2000)& (J2000) &   &   &    &    &    &   &   &    & [mJy] & [mJy]\\
   \noalign{\smallskip}
   \hline
   \noalign{\smallskip}
10:49:58.831& +57:13:31.47&  14.788& 14.01 & 13.36*& 12.49 & 11.98 &  11.23& 10.71& 10.58& 0.90$\pm$0.12& 1.20$\pm$0.04\\
10:50:09.377& +57:04:10.89&  12.916& 12.84 & 12.82*& 11.84*& 11.49*&  11.04& 10.82& 10.70& 0.73$\pm$0.11& 0.93$\pm$0.03\\
10:51:01.477& +57:03:42.03&  14.381& 13.93 & 13.64 & 12.61 & 12.16 &  11.52& 11.12& 11.03& 0.43$\pm$0.06& 0.71$\pm$0.02\\
10:51:30.919& +57:34:39.72&  18.010& 17.11 & 16.37 & 14.69 & 12.79 &  11.04& 10.47& 10.21& 2.41$\pm$0.30& 1.91$\pm$0.07\\
10:50:58.259& +57:25:12.28&  13.164& 13.01 & 12.91*& 11.88*& 11.47 &  10.93& 10.54& 10.46& 0.53$\pm$0.08& 1.25$\pm$0.04\\
10:50:59.130& +57:24:26.44&  14.137& 13.82 & 13.54 & 12.61 & 12.20 &  11.63& 11.23& 11.14& 0.31$\pm$0.06& 0.61$\pm$0.02\\
10:51:59.816& +57:13:12.37&  13.573& 13.33 & 12.89*& 12.14*& 11.72 &  11.16& 10.77& 10.73& 0.91$\pm$0.14& 0.97$\pm$0.03\\
10:52:51.858& +57:27:37.85&  13.071& 13.13 & 12.87*& 12.26 & 11.93 &  11.58& 11.35& 11.27& 0.78$\pm$0.11& 0.58$\pm$0.02\\
10:52:50.223& +57:26:08.61&  12.462& 12.61 & 12.54*& 11.83*& 11.44 &  11.06& 10.82& 10.75& 0.98$\pm$0.12& 0.94$\pm$0.03\\
10:53:10.858& +57:13:55.95&  12.556& 12.48*& 12.28*& 11.62*& 11.14*&  10.69& 10.42& 10.33& 1.38$\pm$0.17& 1.32$\pm$0.04\\
10:53:07.946& +57:18:25.70&  12.893& 12.66 & 12.37*& 11.66*& 11.17*&  10.73& 10.42& 10.33& 1.35$\pm$0.17& 1.34$\pm$0.04\\
10:53:08.928& +57:33:16.41&  13.239& 12.91 & 12.50*& 11.66*& 11.19*&  10.60& 10.21& 10.11& 2.09$\pm$0.26& 1.78$\pm$0.05\\
10:52:01.560& +57:10:46.90&  12.485& 11.99 & 11.83*& 10.97*& 10.38*&   9.46&  9.03&  8.91& 5.20$\pm$0.63& 5.43$\pm$0.15\\
10:51:53.820& +57:19:00.30&  11.288& 11.46*& 11.26*& 10.66*& 10.23*&   9.48&  9.16&  9.11& 4.18$\pm$0.50& 4.06$\pm$0.11\\
10:51:22.040& +57:24:15.30&  12.744& 11.97*& 11.83*& 10.75*& 10.19*&   9.19&  8.65&  8.54& 7.25$\pm$0.87& 7.58$\pm$0.20\\
10:51:14.843& +57:35:25.84&  12.402& 12.21 & 12.28 & 11.29 & 10.71*&  10.25&  9.93&  9.86& 2.01$\pm$0.25& 2.23$\pm$0.07\\
10:49:45.966& +57:03:55.64&  12.247& 11.84 & 11.84 & 10.82 & 10.37*&   9.62&  9.31&  9.17& 4.48$\pm$0.54& 4.10$\pm$0.12\\
10:50:36.049& +57:06:13.27&  11.880& 11.86 & 11.88 & 11.01 & 10.56*&  10.02&  9.71&  9.67& 2.54$\pm$0.31& 2.56$\pm$0.07\\
10:51:10.844& +57:21:40.86&  19.340& 18.31 & 17.64 & 16.01 & 14.16 &  12.40& 11.79& 11.55& 0.65$\pm$0.09& 0.57$\pm$0.02\\
10:52:31.335& +57:15:52.54&  14.061& 14.08 & 13.64 & 13.12 & 12.74 &  12.29& 11.96& 11.90& 0.37$\pm$0.06& 0.31$\pm$0.01\\
10:52:33.483& +57:12:33.94&  14.922& 14.50 & 13.94 & 13.25 & 12.84 &  12.31& 11.98& 11.89& 0.23$\pm$0.06& 0.35$\pm$0.01\\
   \noalign{\smallskip}
   \hline
\end{tabular}
\begin{list}{}{}
\item[$^{\mathrm{*}}$] The star is slightly saturated in this band.
\end{list}
\end{table*}
%


%% file: LHSStab3.tex
%
\renewcommand{\arraystretch}{0.9}
\begin{table*}
\caption[]{LW3 source catalogue: highly reliable detections.}
\label{tab:catalog1}
\tiny
\begin{tabular}{c ccc ccc cc rr cl}
   \hline
   \noalign{\smallskip}
   Source name &    RA   &    DEC  & $\Delta$ & Flux   &  N$_{fr}$ &SNR  & r'  &SNR\_r&\multicolumn{2}{c}{$\Delta$(ISO-opt)}& Prob & Notes\\
               & (J2000) & (J2000) & [arcsec] & [mJy]  &           &    &   [mag]&    &\multicolumn{2}{c}{ [arcsec] }        &      &      \\
   \noalign{\smallskip}
   \hline
   \noalign{\smallskip}
ISO\_LHSS\_J105301+57420 & 10:53:01.124 & +57:42:08.87 & 1.2 & 18.75 $\pm$ 2.25 & 1 & 138 &       &      &  0.6 &  0.9 & $<$0.01 & star \\ 
ISO\_LHSS\_J105349+57070 & 10:53:49.570 & +57:07:07.77 & 1.2 & 11.15 $\pm$ 1.61 & 2 &  58 & 13.59 & 5428 &  0.8 &  0.2 & $<$0.01 & NGC3440 \\ 
ISO\_LHSS\_J105122+57241 & 10:51:22.009 & +57:24:14.36 & 1.2 &  7.27 $\pm$ 0.87 & 4 & 124 & 10.75 & 3619 &  0.2 & -0.3 & $<$0.01 & star \\ 
ISO\_LHSS\_J105035+57332 & 10:50:35.581 & +57:33:22.90 & 1.2 &  7.24 $\pm$ 0.87 & 3 & 151 &       &      &  0.1 & -1.2 & $<$0.01 & star \\ 
ISO\_LHSS\_J105301+57054 & 10:53:01.406 & +57:05:43.64 & 1.2 &  7.10 $\pm$ 0.85 & 4 & 132 & 16.92 & 1551 &  0.5 &  0.5 & $<$0.01 &      \\ 
ISO\_LHSS\_J105201+57104 & 10:52:01.201 & +57:10:43.50 & 1.2 &  5.20 $\pm$ 0.63 & 3 &  85 & 10.97 & 2714 &  0.2 & -0.9 &    0.01 & star \\ 
ISO\_LHSS\_J105336+57380 & 10:53:36.262 & +57:38:00.93 & 1.2 &  5.16 $\pm$ 0.62 & 2 &  61 &       &      & -0.6 &  0.5 & $<$0.01 & star \\ 
ISO\_LHSS\_J105432+57093 & 10:54:32.260 & +57:09:31.23 & 1.2 &  4.80 $\pm$ 0.58 & 2 &  57 & 15.87 & 2171 & -0.4 & -1.3 & $<$0.01 &      \\ 
ISO\_LHSS\_J105445+57274 & 10:54:45.861 & +57:27:47.75 & 1.2 &  4.78 $\pm$ 0.58 & 3 &  65 & 16.05 & 1810 & -0.2 &  0.2 & $<$0.01 &      \\ 
ISO\_LHSS\_J104945+57035 & 10:49:45.959 & +57:03:56.37 & 1.2 &  4.48 $\pm$ 0.54 & 3 &  55 & 10.82 & 3619 & -0.1 &  0.7 & $<$0.01 & star \\ 
ISO\_LHSS\_J105227+57135 & 10:52:27.484 & +57:13:54.67 & 1.2 &  4.43 $\pm$ 0.53 & 4 &  82 & 21.49 &   57 &  0.2 & -0.3 & $<$0.01 &      \\ 
ISO\_LHSS\_J105303+57120 & 10:53:03.798 & +57:12:05.72 & 1.2 &  4.40 $\pm$ 0.53 & 4 &  77 & 19.34 &  310 &  1.0 &  0.1 & $<$0.01 &      \\ 
ISO\_LHSS\_J105052+57350 & 10:50:52.474 & +57:35:06.89 & 1.2 &  4.31 $\pm$ 0.52 & 4 &  66 & 15.31 & 2714 &  0.6 & -0.3 & $<$0.01 &      \\ 
ISO\_LHSS\_J105153+57185 & 10:51:53.683 & +57:18:58.39 & 1.2 &  4.18 $\pm$ 0.50 & 3 &  78 & 10.66 & 3620 & -1.0 &  0.8 & $<$0.01 & star \\ 
ISO\_LHSS\_J105041+57070 & 10:50:41.931 & +57:07:07.73 & 1.2 &  4.11 $\pm$ 0.50 & 4 &  72 & 16.48 &  987 & -0.2 &  1.0 & $<$0.01 &      \\ 
ISO\_LHSS\_J105225+57015 & 10:52:25.876 & +57:01:54.59 & 1.2 &  3.84 $\pm$ 0.47 & 4 &  55 & 16.47 & 1206 &  1.2 &  0.5 & $<$0.01 &      \\ 
ISO\_LHSS\_J105242+57244 & 10:52:42.338 & +57:24:44.69 & 1.2 &  3.81 $\pm$ 0.46 & 3 &  54 & 17.14 & 1810 & -0.6 & -0.3 & $<$0.01 &      \\ 
ISO\_LHSS\_J105128+57350 & 10:51:28.121 & +57:35:02.30 & 1.2 &  3.51 $\pm$ 0.43 & 4 &  60 & 16.70 & 2171 &  0.3 & -0.5 & $<$0.01 &      \\ 
ISO\_LHSS\_J104948+57345 & 10:49:48.882 & +57:34:58.31 & 1.2 &  3.48 $\pm$ 0.42 & 3 &  53 & 17.40 &  905 & -0.1 &  0.1 & $<$0.01 &      \\ 
ISO\_LHSS\_J105404+57401 & 10:54:04.105 & +57:40:19.20 & 1.2 &  3.29 $\pm$ 0.40 & 1 &  38 & 17.99* &  905 &  0.2 & -0.7 & $<$0.01 & star \\ 
ISO\_LHSS\_J105328+57111 & 10:53:28.008 & +57:11:15.54 & 1.2 &  3.23 $\pm$ 0.39 & 4 &  54 & 18.27 &  603 &  0.2 & -0.3 & $<$0.01 &      \\ 
ISO\_LHSS\_J105228+57091 & 10:52:28.678 & +57:09:19.38 & 1.2 &  3.17 $\pm$ 0.38 & 4 &  55 & 19.79 &  241 &  1.2 & -0.4 & $<$0.01 &      \\ 
ISO\_LHSS\_J105318+57214 & 10:53:18.933 & +57:21:40.82 & 1.2 &  3.13 $\pm$ 0.38 & 4 &  48 & 17.21 &  776 &  0.2 &  0.1 & $<$0.01 &      \\ 
ISO\_LHSS\_J105207+57074 & 10:52:07.152 & +57:07:45.35 & 1.2 &  2.79 $\pm$ 0.34 & 3 &  47 & 17.18 &  835 & -0.2 & -0.6 & $<$0.01 &      \\ 
ISO\_LHSS\_J105256+57082 & 10:52:56.829 & +57:08:24.64 & 1.2 &  2.69 $\pm$ 0.33 & 4 &  36 & 16.88 &  987 &  0.2 & -1.2 & $<$0.01 &      \\ 
ISO\_LHSS\_J105421+57254 & 10:54:21.270 & +57:25:44.84 & 1.2 &  2.64 $\pm$ 0.32 & 3 &  41 & 18.63 &  835 &  0.8 &  0.3 & $<$0.01 &      \\ 
ISO\_LHSS\_J105242+57191 & 10:52:42.294 & +57:19:14.50 & 1.2 &  2.60 $\pm$ 0.32 & 5 &  50 & 17.55 & 2171 & -1.1 & -0.3 & $<$0.01 &      \\ 
ISO\_LHSS\_J105035+57061 & 10:50:35.929 & +57:06:13.08 & 1.2 &  2.55 $\pm$ 0.31 & 4 &  37 & 11.01 & 2714 & -1.0 & -0.2 & $<$0.01 & star \\ 
ISO\_LHSS\_J105231+57175 & 10:52:31.674 & +57:17:51.98 & 1.2 &  2.46 $\pm$ 0.30 & 3 &  33 & 22.99 &   24 &  1.1 & -0.1 &    0.01 &      \\ 
ISO\_LHSS\_J105143+57293 & 10:51:43.660 & +57:29:38.89 & 1.3 &  2.41 $\pm$ 0.30 & 3 &  28 & 16.39 & 1810 & -0.9 &  1.8 & $<$0.01 &      \\ 
ISO\_LHSS\_J105130+57343 & 10:51:30.985 & +57:34:39.12 & 1.2 &  2.41 $\pm$ 0.30 & 3 &  39 & 14.69* &  517 & -0.6 & -0.8 & $<$0.01 & star \\ 
ISO\_LHSS\_J105252+57290 & 10:52:52.753 & +57:29:01.68 & 1.2 &  2.21 $\pm$ 0.27 & 4 &  44 & 17.32 & 1357 & -0.0 &  1.9 & $<$0.01 &      \\ 
ISO\_LHSS\_J105309+57331 & 10:53:09.093 & +57:33:17.14 & 1.4 &  2.09 $\pm$ 0.27 & 2 &  24 & 11.66 & 2171 &  1.2 &  0.6 & $<$0.01 & star \\ 
ISO\_LHSS\_J104948+57382 & 10:49:48.651 & +57:38:21.10 & 1.3 &  2.01 $\pm$ 0.25 & 2 &  29 &       &      &  0.4 &  0.2 & $<$0.01 & star \\ 
ISO\_LHSS\_J105114+57352 & 10:51:14.832 & +57:35:25.47 & 1.4 &  2.01 $\pm$ 0.25 & 4 &  25 & 11.29 & 2714 & -0.1 & -0.4 & $<$0.01 & star \\ 
ISO\_LHSS\_J105359+56581 & 10:53:59.546 & +56:58:17.31 & 1.2 &  1.98 $\pm$ 0.25 & 4 &  31 & 18.12 &  517 &  0.4 &  0.3 & $<$0.01 &      \\ 
ISO\_LHSS\_J105141+57450 & 10:51:41.422 & +57:45:07.02 & 2.0 &  1.93 $\pm$ 0.35 & 1 &   7 & 24.89 &    4 &  1.3 & -1.5 &    0.11 &      \\ 
ISO\_LHSS\_J105113+57142 & 10:51:13.381 & +57:14:26.20 & 1.2 &  1.92 $\pm$ 0.24 & 3 &  37 & 20.10 &  184 &  0.5 & -0.6 & $<$0.01 &      \\ 
ISO\_LHSS\_J105426+57364 & 10:54:26.407 & +57:36:49.23 & 1.2 &  1.84 $\pm$ 0.23 & 3 &  33 & 18.08 &  517 &  1.2 & -0.0 & $<$0.01 &      \\ 
ISO\_LHSS\_J105002+57472 & 10:50:02.791 & +57:47:20.91 & 1.6 &  1.79 $\pm$ 0.25 & 1 &  15 & 20.43 &  149 & -0.3 &  0.8 & $<$0.01 & star \\ 
ISO\_LHSS\_J105320+57143 & 10:53:20.790 & +57:14:32.51 & 1.4 &  1.78 $\pm$ 0.23 & 3 &  21 & 15.66 & 1810 & -1.1 & -0.7 & $<$0.01 &      \\ 
ISO\_LHSS\_J105003+57323 & 10:50:03.885 & +57:32:37.79 & 1.4 &  1.65 $\pm$ 0.21 & 3 &  23 &       &      &  0.3 & -0.3 & $<$0.01 & star \\ 
ISO\_LHSS\_J104958+57355 & 10:49:58.700 & +57:35:52.66 & 1.2 &  1.65 $\pm$ 0.20 & 3 &  31 &       &      &  1.3 & -0.6 & $<$0.01 & star \\ 
ISO\_LHSS\_J105100+57411 & 10:51:00.385 & +57:41:15.22 & 1.2 &  1.61 $\pm$ 0.20 & 3 &  33 & 16.13 &  776 & -0.2 &  0.2 & $<$0.01 &      \\ 
ISO\_LHSS\_J105252+57075 & 10:52:52.690 & +57:07:53.17 & 1.5 &  1.57 $\pm$ 0.20 & 2 &  20 & 17.46 & 1086 & -0.5 & -0.7 & $<$0.01 &      \\ 
ISO\_LHSS\_J105434+57202 & 10:54:34.841 & +57:20:25.24 & 1.5 &  1.57 $\pm$ 0.20 & 4 &  20 & 22.88 &   21 &  0.9 &  1.6 &    0.02 &      \\ 
ISO\_LHSS\_J105243+57404 & 10:52:43.469 & +57:40:40.13 & 1.3 &  1.53 $\pm$ 0.19 & 3 &  26 & 16.24* & 1357 &  0.0 &  0.9 & $<$0.01 & star \\ 
ISO\_LHSS\_J105216+57353 & 10:52:16.501 & +57:35:30.13 & 1.3 &  1.51 $\pm$ 0.19 & 3 &  26 & 18.36 &  776 & -0.9 & -0.0 & $<$0.01 &      \\ 
ISO\_LHSS\_J105134+57335 & 10:51:34.343 & +57:33:59.89 & 1.4 &  1.50 $\pm$ 0.19 & 3 &  24 & 16.70 & 1357 & -0.8 & -0.1 & $<$0.01 &      \\ 
ISO\_LHSS\_J105458+57282 & 10:54:58.477 & +57:28:28.13 & 1.3 &  1.47 $\pm$ 0.18 & 3 &  26 & 22.46 &   26 & -0.5 & -0.4 & $<$0.01 &      \\ 
ISO\_LHSS\_J105412+57090 & 10:54:12.220 & +57:09:00.20 & 1.2 &  1.47 $\pm$ 0.18 & 7 &  33 & 15.99* &  542 & -1.8 & -2.1 & $<$0.01 & star \\ 
ISO\_LHSS\_J105314+57413 & 10:53:14.714 & +57:41:33.53 & 1.7 &  1.45 $\pm$ 0.21 & 1 &  13 & 18.41 &  157 & -0.9 & -4.5 & $<$0.01 &      \\ 
ISO\_LHSS\_J105019+57281 & 10:50:19.757 & +57:28:13.17 & 1.4 &  1.42 $\pm$ 0.18 & 3 &  21 & 19.22 &  434 &  0.6 &  0.2 & $<$0.01 &      \\ 
ISO\_LHSS\_J105341+57191 & 10:53:41.034 & +57:19:19.77 & 1.4 &  1.40 $\pm$ 0.18 & 4 &  24 & 17.61 &  776 &  1.5 &  2.0 & $<$0.01 &      \\ 
ISO\_LHSS\_J105351+57072 & 10:53:51.273 & +57:07:26.98 & 1.5 &  1.39 $\pm$ 0.18 & 3 &  17 & 17.21 &  835 &  0.3 & -0.7 & $<$0.01 & NGC3440 (b) \\ 
ISO\_LHSS\_J105310+57135 & 10:53:10.719 & +57:13:56.50 & 1.4 &  1.38 $\pm$ 0.17 & 2 &  24 & 11.62 & 2171 & -0.8 &  0.3 & $<$0.01 & star \\ 
ISO\_LHSS\_J105307+57182 & 10:53:07.906 & +57:18:27.11 & 1.4 &  1.35 $\pm$ 0.17 & 3 &  22 & 11.66 & 2171 & -0.0 &  1.2 & $<$0.01 & star \\ 
ISO\_LHSS\_J105425+57193 & 10:54:25.653 & +57:19:37.60 & 1.4 &  1.34 $\pm$ 0.17 & 4 &  22 & 21.07 &   89 & -0.2 &  0.3 & $<$0.01 &      \\ 
ISO\_LHSS\_J104956+57144 & 10:49:56.045 & +57:14:41.42 & 1.4 &  1.34 $\pm$ 0.17 & 4 &  22 & 17.75 &  494 & -0.4 &  1.0 & $<$0.01 &      \\ 
ISO\_LHSS\_J105156+56583 & 10:51:56.210 & +56:58:33.10 & 1.9 &  1.29 $\pm$ 0.21 & 1 &   8 & 24.65 &    3 & -1.8 &  2.2 &    0.18 &      \\ 
ISO\_LHSS\_J105334+57130 & 10:53:34.138 & +57:13:01.47 & 1.3 &  1.29 $\pm$ 0.16 & 4 &  26 & 19.20 &  418 & -0.5 &  0.2 & $<$0.01 &      \\ 
ISO\_LHSS\_J105326+57140 & 10:53:26.796 & +57:14:05.70 & 1.5 &  1.29 $\pm$ 0.17 & 4 &  19 & 19.34 &  339 &  0.9 &  0.1 & $<$0.01 & pair \\ 
ISO\_LHSS\_J105225+57113 & 10:52:25.649 & +57:11:30.57 & 1.7 &  1.26 $\pm$ 0.18 & 3 &  13 & 17.73 & 1086 &  0.5 & -1.0 & $<$0.01 &      \\ 
ISO\_LHSS\_J105427+57075 & 10:54:27.704 & +57:07:59.57 & 1.4 &  1.26 $\pm$ 0.16 & 3 &  21 & 18.25 &  517 & -0.1 & -0.5 & $<$0.01 &      \\ 
ISO\_LHSS\_J105336+57014 & 10:53:36.438 & +57:01:47.06 & 1.4 &  1.24 $\pm$ 0.16 & 4 &  23 & 17.76 &  679 &  1.0 &  0.1 & $<$0.01 &      \\ 
ISO\_LHSS\_J105200+57180 & 10:52:00.264 & +57:18:04.56 & 1.4 &  1.23 $\pm$ 0.16 & 6 &  25 & 24.01 &   10 & -0.8 & -1.4 &    0.04 &      \\ 
ISO\_LHSS\_J105058+57451 & 10:50:58.682 & +57:45:12.02 & 1.5 &  1.23 $\pm$ 0.16 & 3 &  20 & 17.20 &  905 &  0.4 & -0.5 & $<$0.01 &      \\ 
ISO\_LHSS\_J105330+57392 & 10:53:30.905 & +57:39:21.51 & 1.5 &  1.22 $\pm$ 0.16 & 3 &  19 & 21.42 &   54 & -0.4 &  1.0 & $<$0.01 &      \\ 
ISO\_LHSS\_J105324+57123 & 10:53:24.730 & +57:12:31.92 & 1.7 &  1.22 $\pm$ 0.18 & 2 &  12 & 17.28 & 1357 &  0.6 &  0.8 & $<$0.01 &      \\ 
ISO\_LHSS\_J105315+57412 & 10:53:15.970 & +57:41:25.33 & 1.8 &  1.19 $\pm$ 0.18 & 1 &  11 & 19.07 &  105 &  1.2 & -6.0 &    0.01 &      \\ 
ISO\_LHSS\_J105150+57390 & 10:51:50.273 & +57:39:06.89 & 1.4 &  1.19 $\pm$ 0.15 & 3 &  23 & 17.40 &  776 & -1.8 &  0.8 & $<$0.01 &      \\ 
ISO\_LHSS\_J105411+57101 & 10:54:11.016 & +57:10:16.14 & 1.5 &  1.18 $\pm$ 0.15 & 3 &  19 & 17.33 &  835 & -1.8 & -0.2 & $<$0.01 &      \\ 
ISO\_LHSS\_J105347+57013 & 10:53:47.780 & +57:01:36.27 & 1.7 &  1.17 $\pm$ 0.16 & 2 &  13 & 21.83 &   49 &  1.5 &  0.3 &    0.01 &      \\ 
ISO\_LHSS\_J105414+57124 & 10:54:14.777 & +57:12:42.85 & 1.4 &  1.16 $\pm$ 0.15 & 4 &  21 & 19.20 &  472 &  0.1 &  0.5 & $<$0.01 & star \\ 
ISO\_LHSS\_J105141+57150 & 10:51:41.887 & +57:15:04.33 & 1.5 &  1.15 $\pm$ 0.15 & 2 &  18 & 18.58 &  679 & -1.1 &  2.5 & $<$0.01 & pair \\ 
ISO\_LHSS\_J105239+57243 & 10:52:39.408 & +57:24:31.33 & 1.5 &  1.15 $\pm$ 0.15 & 3 &  19 & 17.73* & 2171 & -1.5 & -0.3 & $<$0.01 & star \\ 
ISO\_LHSS\_J105213+57160 & 10:52:13.539 & +57:16:04.74 & 1.4 &  1.15 $\pm$ 0.15 & 3 &  22 & 22.52 &   27 &  1.2 & -0.5 &    0.01 &      \\ 
ISO\_LHSS\_J105134+57415 & 10:51:34.644 & +57:41:54.56 & 1.4 &  1.12 $\pm$ 0.14 & 3 &  21 & 17.42 &  679 &  0.1 &  0.2 & $<$0.01 &      \\ 
ISO\_LHSS\_J105039+57452 & 10:50:39.598 & +57:45:27.99 & 1.6 &  1.11 $\pm$ 0.15 & 2 &  14 & 15.78 & 2171 &  0.6 &  0.5 & $<$0.01 &      \\ 
ISO\_LHSS\_J104949+57375 & 10:49:49.977 & +57:37:57.67 & 1.5 &  1.10 $\pm$ 0.15 & 3 &  17 & 21.30 &   80 & -0.1 &  0.3 & $<$0.01 &      \\ 
ISO\_LHSS\_J105438+57225 & 10:54:38.419 & +57:22:55.78 & 1.7 &  1.08 $\pm$ 0.15 & 3 &  13 & 22.15 &   26 &  1.1 & -0.1 & $<$0.01 &      \\ 
ISO\_LHSS\_J105152+57090 & 10:51:52.046 & +57:09:08.81 & 1.6 &  1.07 $\pm$ 0.15 & 2 &  15 & 19.21 &  241 & -0.3 & -0.4 & $<$0.01 &      \\ 
ISO\_LHSS\_J105427+57100 & 10:54:27.012 & +57:10:03.97 & 1.5 &  1.07 $\pm$ 0.14 & 3 &  17 & 21.72 &   41 & -0.9 & -0.4 & $<$0.01 &      \\ 
ISO\_LHSS\_J105343+57253 & 10:53:43.348 & +57:25:30.29 & 1.6 &  1.06 $\pm$ 0.14 & 4 &  16 & 20.64 &  122 &  1.2 &  0.4 & $<$0.01 &      \\ 
ISO\_LHSS\_J105257+57151 & 10:52:57.678 & +57:15:15.93 & 1.4 &  1.05 $\pm$ 0.13 & 6 &  25 & 18.03 &  724 &  0.1 &  0.2 & $<$0.01 &      \\ 
ISO\_LHSS\_J105354+57052 & 10:53:54.174 & +57:05:27.13 & 1.6 &  1.04 $\pm$ 0.14 & 3 &  16 & 19.31 &  319 & -0.9 &  1.0 & $<$0.01 &      \\ 
ISO\_LHSS\_J105404+57203 & 10:54:04.211 & +57:20:35.39 & 1.6 &  1.03 $\pm$ 0.14 & 2 &  15 & 19.58 &  418 & -0.7 & -1.0 & $<$0.01 &      \\ 
ISO\_LHSS\_J105345+56552 & 10:53:45.908 & +56:55:27.72 & 2.0 &  1.00 $\pm$ 0.17 & 1 &   8 & 23.34 &    8 &  1.5 &  2.9 &    0.10 &      \\ 
ISO\_LHSS\_J105319+57124 & 10:53:19.545 & +57:12:44.21 & 1.6 &  0.99 $\pm$ 0.14 & 2 &  14 & 22.83 &   19 &  0.1 & -1.8 &    0.02 &      \\ 
ISO\_LHSS\_J105126+57215 & 10:51:26.646 & +57:21:59.72 & 1.6 &  0.98 $\pm$ 0.13 & 5 &  15 & 19.47 &  494 & -0.9 & -0.7 & $<$0.01 &      \\ 
ISO\_LHSS\_J105255+57195 & 10:52:55.481 & +57:19:51.99 & 1.6 &  0.98 $\pm$ 0.13 & 2 &  15 & 24.30 &    6 &  1.3 &  2.3 &    0.13 &      \\ 
ISO\_LHSS\_J105250+57260 & 10:52:50.255 & +57:26:08.20 & 1.5 &  0.98 $\pm$ 0.13 & 4 &  19 & 11.83 & 2171 &  0.1 & -0.3 & $<$0.01 & star \\ 
ISO\_LHSS\_J105316+57355 & 10:53:16.641 & +57:35:51.37 & 1.8 &  0.98 $\pm$ 0.15 & 2 &  11 & 19.01 &  603 & -0.7 &  0.5 & $<$0.01 & star \\ 
ISO\_LHSS\_J105308+57132 & 10:53:08.229 & +57:13:22.52 & 1.7 &  0.96 $\pm$ 0.14 & 3 &  11 & 19.57 &  205 & -1.0 & -0.1 & $<$0.01 &      \\ 
ISO\_LHSS\_J105413+57263 & 10:54:13.085 & +57:26:31.75 & 1.6 &  0.96 $\pm$ 0.13 & 3 &  15 & 15.60* &  417 &  1.0 &  0.7 & $<$0.01 & star \\ 
ISO\_LHSS\_J104952+57082 & 10:49:52.837 & +57:08:20.30 & 1.6 &  0.96 $\pm$ 0.13 & 4 &  14 & 20.13 &  201 & -0.7 & -1.0 & $<$0.01 &      \\ 
ISO\_LHSS\_J105103+57431 & 10:51:03.311 & +57:43:16.93 & 1.5 &  0.96 $\pm$ 0.13 & 4 &  18 & 19.37 &  310 & -0.4 & -0.0 & $<$0.01 &      \\ 
ISO\_LHSS\_J105331+57340 & 10:53:31.857 & +57:34:09.57 & 1.6 &  0.95 $\pm$ 0.13 & 4 &  15 & 15.63* &  684 &  0.1 & -0.7 & $<$0.01 & star \\ 
ISO\_LHSS\_J105340+57045 & 10:53:40.331 & +57:04:55.50 & 1.7 &  0.95 $\pm$ 0.13 & 3 &  13 & 18.34 &  494 &  1.8 &  0.3 & $<$0.01 &      \\ 
ISO\_LHSS\_J105018+57462 & 10:50:18.073 & +57:46:23.65 & 1.6 &  0.94 $\pm$ 0.13 & 2 &  14 & 20.28 &  181 & -0.8 & -0.8 & $<$0.01 &      \\ 
   \noalign{\smallskip}
   \hline
\end{tabular}
\end{table*}

\newpage

\setcounter{table}{2}
\begin{table*}
\caption[]{Continue.}
\label{tab:catalog1}
\tiny
\begin{tabular}{c ccc ccc cc rr cl}
   \hline
   \noalign{\smallskip}
   Source name &    RA   &    DEC  & $\Delta$ & Flux   &  N$_{fr}$ &SNR  & r'  &SNR\_r&\multicolumn{2}{c}{$\Delta$(ISO-opt)}& Prob & Notes\\
               & (J2000) & (J2000) & [arcsec] & [mJy]  &           &    &   [mag]&    &\multicolumn{2}{c}{ [arcsec] }        &      &      \\
   \noalign{\smallskip}
   \hline
   \noalign{\smallskip}
ISO\_LHSS\_J105456+57301 & 10:54:56.268 & +57:30:18.11 & 1.6 &  0.93 $\pm$ 0.13 & 3 &  14 & 20.82 &   62 & -0.1 &  0.4 & $<$0.01 &      \\ 
ISO\_LHSS\_J105335+57195 & 10:53:35.801 & +57:19:54.98 & 1.7 &  0.92 $\pm$ 0.13 & 3 &  13 & 15.84* &  639 & -0.4 & -0.9 & $<$0.01 & star \\ 
ISO\_LHSS\_J104905+57130 & 10:49:05.684 & +57:13:08.08 & 1.6 &  0.92 $\pm$ 0.13 & 2 &  14 & 19.40 &  265 & -2.4 & -0.4 & $<$0.01 &      \\ 
ISO\_LHSS\_J105327+57093 & 10:53:27.484 & +57:09:31.15 & 1.6 &  0.92 $\pm$ 0.13 & 4 &  14 & 20.14 &  187 &  0.0 & -0.7 & $<$0.01 &      \\ 
ISO\_LHSS\_J105409+57101 & 10:54:09.965 & +57:10:14.08 & 1.8 &  0.91 $\pm$ 0.14 & 2 &  10 &       &      &      &      &         & bridge\\ 
ISO\_LHSS\_J105159+57131 & 10:51:59.769 & +57:13:11.21 & 1.9 &  0.91 $\pm$ 0.15 & 3 &   9 & 12.14 & 1809 & -0.5 &  2.1 & $<$0.01 & star \\ 
ISO\_LHSS\_J105246+57174 & 10:52:46.278 & +57:17:47.67 & 1.6 &  0.90 $\pm$ 0.12 & 3 &  14 & 20.65 &   74 & -0.5 & -0.7 & $<$0.01 &      \\ 
ISO\_LHSS\_J105054+57353 & 10:50:54.946 & +57:35:30.81 & 1.6 &  0.90 $\pm$ 0.12 & 2 &  14 & 22.51 &   30 & -0.3 & -1.2 &    0.01 &      \\ 
ISO\_LHSS\_J104958+57133 & 10:49:58.982 & +57:13:32.36 & 1.7 &  0.90 $\pm$ 0.13 & 3 &  12 & 12.61 & 1551 &  0.7 &  0.5 & $<$0.01 & star \\ 
ISO\_LHSS\_J105048+57430 & 10:50:48.351 & +57:43:04.08 & 1.7 &  0.88 $\pm$ 0.12 & 2 &  13 & 21.31 &   65 &  0.5 & -0.2 & $<$0.01 & pair \\ 
ISO\_LHSS\_J105107+57455 & 10:51:07.665 & +57:45:50.41 & 1.9 &  0.88 $\pm$ 0.14 & 1 &   9 & 24.82 &    3 & -2.6 & -2.9 &    0.34 &      \\ 
ISO\_LHSS\_J105045+57082 & 10:50:45.692 & +57:08:22.83 & 1.5 &  0.88 $\pm$ 0.11 & 5 &  19 & 21.72 &   67 &  0.4 & -0.7 & $<$0.01 & pair \\ 
ISO\_LHSS\_J105351+57352 & 10:53:51.119 & +57:35:27.38 & 1.6 &  0.87 $\pm$ 0.12 & 6 &  15 & 21.99 &   44 &  0.2 & -0.6 & $<$0.01 &      \\ 
ISO\_LHSS\_J105217+57212 & 10:52:17.765 & +57:21:27.56 & 1.6 &  0.87 $\pm$ 0.12 & 3 &  16 & 21.79 &   64 &  1.2 &  0.2 & $<$0.01 &      \\ 
ISO\_LHSS\_J105232+56571 & 10:52:32.307 & +56:57:17.36 & 1.8 &  0.86 $\pm$ 0.13 & 1 &  10 & 20.65 &  139 &  2.2 &  0.8 &    0.01 &      \\ 
ISO\_LHSS\_J105125+57254 & 10:51:25.078 & +57:25:43.36 & 1.6 &  0.84 $\pm$ 0.12 & 3 &  15 & 21.92 &   54 & -3.1 & -0.9 &    0.03 &      \\ 
ISO\_LHSS\_J105307+57145 & 10:53:07.885 & +57:14:59.05 & 1.7 &  0.84 $\pm$ 0.12 & 3 &  13 & 21.68 &   34 &  0.5 & -0.8 & $<$0.01 &      \\ 
ISO\_LHSS\_J105220+57340 & 10:52:20.826 & +57:34:07.93 & 1.6 &  0.82 $\pm$ 0.11 & 3 &  14 & 22.91 &   19 &  0.7 & -0.7 &    0.01 &      \\ 
ISO\_LHSS\_J105301+57343 & 10:53:01.436 & +57:34:30.68 & 1.7 &  0.81 $\pm$ 0.12 & 2 &  12 & 21.97 &   44 &  0.5 &  0.3 & $<$0.01 &      \\ 
ISO\_LHSS\_J105234+57264 & 10:52:34.944 & +57:26:45.31 & 1.7 &  0.81 $\pm$ 0.11 & 3 &  13 & 21.17 &  136 &  1.0 &  1.8 &    0.01 & star \\ 
ISO\_LHSS\_J105029+57103 & 10:50:29.421 & +57:10:37.37 & 1.7 &  0.80 $\pm$ 0.11 & 3 &  13 & 18.63 &  452 & -2.0 &  1.0 & $<$0.01 &      \\ 
ISO\_LHSS\_J105211+57433 & 10:52:11.429 & +57:43:31.94 & 2.0 &  0.80 $\pm$ 0.14 & 1 &   7 & 23.39 &   15 & -1.3 & -0.1 &    0.02 &      \\ 
ISO\_LHSS\_J105312+57201 & 10:53:12.015 & +57:20:13.74 & 1.8 &  0.79 $\pm$ 0.12 & 2 &  10 & 23.66 &    8 &  3.2 & -1.5 &    0.14 &  ?   \\ 
ISO\_LHSS\_J104903+57063 & 10:49:03.120 & +57:06:38.58 & 1.6 &  0.79 $\pm$ 0.11 & 3 &  15 & 19.90 &  247 & -0.8 &  0.0 & $<$0.01 &      \\ 
ISO\_LHSS\_J105251+57273 & 10:52:51.855 & +57:27:36.55 & 1.7 &  0.78 $\pm$ 0.11 & 3 &  13 & 12.26 & 1809 &  0.1 & -2.1 & $<$0.01 & star \\ 
ISO\_LHSS\_J105149+57330 & 10:51:49.772 & +57:33:02.25 & 1.8 &  0.77 $\pm$ 0.12 & 2 &  10 &       &      &      &      &         &      \\ 
ISO\_LHSS\_J105157+56581 & 10:51:57.352 & +56:58:10.91 & 1.9 &  0.77 $\pm$ 0.13 & 1 &   8 & 23.66 &   10 &  0.7 & -0.3 &    0.01 &      \\ 
ISO\_LHSS\_J105237+57143 & 10:52:37.482 & +57:14:33.62 & 1.7 &  0.77 $\pm$ 0.11 & 3 &  12 & 21.17 &   82 & -0.6 &  0.2 & $<$0.01 &      \\ 
ISO\_LHSS\_J105047+57014 & 10:50:47.681 & +57:01:45.89 & 1.9 &  0.77 $\pm$ 0.13 & 2 &   8 &       &      & -0.3 & -2.4 & $<$0.01 & star \\ 
ISO\_LHSS\_J105440+57140 & 10:54:40.653 & +57:14:03.76 & 1.9 &  0.76 $\pm$ 0.12 & 3 &   9 & 21.67 &   42 &  0.2 & -1.2 & $<$0.01 &      \\ 
ISO\_LHSS\_J105304+57053 & 10:53:04.347 & +57:05:30.66 & 1.8 &  0.75 $\pm$ 0.12 & 4 &  10 & 21.25 &   57 & -1.7 &  0.9 &    0.01 &      \\ 
ISO\_LHSS\_J105034+57410 & 10:50:34.398 & +57:41:05.10 & 1.7 &  0.74 $\pm$ 0.11 & 3 &  12 & 15.92* & 1206 & -2.1 &  0.6 & $<$0.01 & star \\ 
ISO\_LHSS\_J105025+57331 & 10:50:25.107 & +57:33:12.33 & 1.7 &  0.74 $\pm$ 0.10 & 4 &  13 & 20.48 &  151 & -1.9 & -1.2 & $<$0.01 &      \\ 
ISO\_LHSS\_J105009+57041 & 10:50:09.291 & +57:04:11.37 & 1.9 &  0.73 $\pm$ 0.12 & 4 &   9 & 11.84 & 2171 & -0.7 &  0.4 & $<$0.01 & star \\ 
ISO\_LHSS\_J105306+57280 & 10:53:06.617 & +57:28:06.76 & 1.7 &  0.73 $\pm$ 0.10 & 3 &  12 & 22.02 &   44 & -0.5 &  0.1 & $<$0.01 &      \\ 
ISO\_LHSS\_J104940+57161 & 10:49:40.591 & +57:16:10.53 & 2.0 &  0.73 $\pm$ 0.13 & 2 &   7 & 22.86 &   15 &  0.4 &  0.4 & $<$0.01 &      \\ 
ISO\_LHSS\_J105404+57332 & 10:54:04.537 & +57:33:28.29 & 1.9 &  0.72 $\pm$ 0.12 & 2 &   8 & 21.25 &   80 & -0.2 & -0.6 & $<$0.01 &      \\ 
ISO\_LHSS\_J105410+57090 & 10:54:10.488 & +57:09:09.30 & 1.7 &  0.72 $\pm$ 0.10 & 4 &  12 & 19.16 &   65 &  0.8 & -1.1 & $<$0.01 &      \\ 
ISO\_LHSS\_J105019+57434 & 10:50:19.856 & +57:43:46.75 & 1.8 &  0.71 $\pm$ 0.10 & 3 &  11 & 18.77 &  571 & -2.1 & -1.0 & $<$0.01 &      \\ 
ISO\_LHSS\_J105231+57064 & 10:52:31.963 & +57:06:49.42 & 1.8 &  0.69 $\pm$ 0.10 & 3 &  11 & 19.33 &  241 &  1.3 & -1.6 & $<$0.01 &      \\ 
ISO\_LHSS\_J105435+57171 & 10:54:35.735 & +57:17:10.01 & 1.8 &  0.69 $\pm$ 0.10 & 2 &  11 & 20.96 &   91 & -0.6 &  1.0 & $<$0.01 & pair \\ 
ISO\_LHSS\_J105031+57082 & 10:50:31.124 & +57:08:29.48 & 1.8 &  0.69 $\pm$ 0.10 & 2 &  11 & 21.78 &   43 &  0.7 &  1.3 &    0.01 &      \\ 
ISO\_LHSS\_J105232+57051 & 10:52:32.197 & +57:05:12.09 & 1.9 &  0.68 $\pm$ 0.11 & 3 &   9 & 19.80 &  265 & -1.5 & -0.6 & $<$0.01 &      \\ 
ISO\_LHSS\_J105151+56583 & 10:51:51.742 & +56:58:30.90 & 2.2 &  0.68 $\pm$ 0.14 & 1 &   6 & 20.41 &  145 & -2.0 &  1.8 &    0.01 &      \\ 
ISO\_LHSS\_J105439+57200 & 10:54:39.056 & +57:20:08.54 & 1.8 &  0.68 $\pm$ 0.10 & 4 &  10 & 18.68* &  835 & -0.9 & -0.3 & $<$0.01 & star \\ 
ISO\_LHSS\_J104953+57381 & 10:49:53.936 & +57:38:14.30 & 1.9 &  0.68 $\pm$ 0.11 & 3 &   9 & 21.87 &   48 &  1.2 & -1.6 &    0.01 &      \\ 
ISO\_LHSS\_J105227+57141 & 10:52:27.561 & +57:14:13.43 & 2.1 &  0.67 $\pm$ 0.13 & 2 &   6 & 19.85 &  247 &  1.1 & -1.9 & $<$0.01 &      \\ 
ISO\_LHSS\_J104944+57342 & 10:49:44.938 & +57:34:23.77 & 1.7 &  0.67 $\pm$ 0.10 & 3 &  12 & 18.63 &  339 & -1.9 &  0.1 & $<$0.01 &      \\ 
ISO\_LHSS\_J105115+57355 & 10:51:15.168 & +57:35:51.97 & 1.7 &  0.67 $\pm$ 0.10 & 3 &  11 & 21.63 &   22 &  1.7 & -0.5 &    0.01 &      \\ 
ISO\_LHSS\_J104911+57140 & 10:49:11.598 & +57:14:09.31 & 1.9 &  0.67 $\pm$ 0.11 & 2 &   8 & 22.26 &   34 &  0.8 &  0.2 & $<$0.01 &      \\ 
ISO\_LHSS\_J105154+57300 & 10:51:54.199 & +57:30:08.82 & 2.0 &  0.67 $\pm$ 0.11 & 2 &   8 & 24.73 &    5 &  1.1 &  0.1 &    0.03 &      \\ 
ISO\_LHSS\_J105013+57113 & 10:50:13.132 & +57:11:39.57 & 2.0 &  0.66 $\pm$ 0.11 & 2 &   8 & 20.44 &  136 & -1.5 &  1.4 & $<$0.01 &      \\ 
ISO\_LHSS\_J105015+57414 & 10:50:15.535 & +57:41:40.01 & 1.9 &  0.66 $\pm$ 0.10 & 3 &   9 & 22.73 &   18 &  0.4 &  2.2 &    0.03 & star \\ 
ISO\_LHSS\_J105257+57012 & 10:52:57.671 & +57:01:20.26 & 2.1 &  0.66 $\pm$ 0.13 & 2 &   6 & 22.40 &   27 &  1.2 & -0.3 &    0.01 &      \\ 
ISO\_LHSS\_J105110+57214 & 10:51:10.730 & +57:21:41.85 & 1.8 &  0.65 $\pm$ 0.10 & 3 &  10 & 16.01* &  231 & -0.8 &  1.1 & $<$0.01 & star \\ 
ISO\_LHSS\_J105135+57274 & 10:51:35.728 & +57:27:40.86 & 1.8 &  0.65 $\pm$ 0.10 & 4 &  10 & 22.73 &   34 &  1.7 &  1.6 &    0.03 &      \\ 
ISO\_LHSS\_J105148+57380 & 10:51:48.186 & +57:38:05.49 & 1.7 &  0.65 $\pm$ 0.09 & 4 &  12 & 21.62 &   50 & -1.4 & -1.6 &    0.01 &      \\ 
ISO\_LHSS\_J105136+57300 & 10:51:36.160 & +57:30:00.37 & 1.9 &  0.64 $\pm$ 0.10 & 3 &   9 & 21.82 &   58 & -0.1 &  0.5 & $<$0.01 &      \\ 
ISO\_LHSS\_J105242+57004 & 10:52:42.524 & +57:00:47.96 & 1.9 &  0.64 $\pm$ 0.10 & 3 &   9 & 22.66 &   24 &  2.9 & -0.6 &    0.05 &      \\ 
ISO\_LHSS\_J105256+56594 & 10:52:56.078 & +56:59:46.91 & 1.9 &  0.64 $\pm$ 0.11 & 3 &   8 & 18.27 &  494 & -1.8 & -0.5 & $<$0.01 &      \\ 
ISO\_LHSS\_J105350+57091 & 10:53:50.449 & +57:09:15.24 & 1.6 &  0.64 $\pm$ 0.09 & 5 &  15 & 19.64 &  271 & -2.0 & -0.4 & $<$0.01 &      \\ 
ISO\_LHSS\_J104938+57130 & 10:49:38.243 & +57:13:07.84 & 1.8 &  0.63 $\pm$ 0.09 & 4 &  11 & 19.96 &  226 &  0.5 & -0.1 & $<$0.01 &      \\ 
ISO\_LHSS\_J105215+57263 & 10:52:15.103 & +57:26:34.89 & 1.8 &  0.63 $\pm$ 0.10 & 6 &  10 & 19.31 &  418 &  0.6 & -0.3 & $<$0.01 &      \\ 
ISO\_LHSS\_J105417+57002 & 10:54:17.278 & +57:00:25.17 & 1.7 &  0.63 $\pm$ 0.09 & 4 &  12 & 19.38 &  310 & -0.8 & -0.7 & $<$0.01 &      \\ 
ISO\_LHSS\_J105024+57141 & 10:50:24.694 & +57:14:13.53 & 1.9 &  0.63 $\pm$ 0.10 & 3 &   9 & 18.32 &  571 & -0.6 & -1.2 & $<$0.01 &      \\ 
ISO\_LHSS\_J105103+57203 & 10:51:03.538 & +57:20:37.94 & 1.6 &  0.62 $\pm$ 0.09 & 5 &  14 & 19.63 &  434 &  0.7 &  0.7 & $<$0.01 &      \\ 
ISO\_LHSS\_J105409+57012 & 10:54:09.305 & +57:01:23.37 & 1.8 &  0.62 $\pm$ 0.10 & 3 &  10 & 20.35 &   88 &  1.5 & -0.1 & $<$0.01 &      \\ 
ISO\_LHSS\_J105020+57424 & 10:50:20.566 & +57:42:40.42 & 1.8 &  0.62 $\pm$ 0.10 & 3 &  10 & 21.52 &   72 &  0.7 &  0.1 & $<$0.01 & star \\ 
ISO\_LHSS\_J105242+57421 & 10:52:42.177 & +57:42:18.21 & 2.1 &  0.62 $\pm$ 0.12 & 2 &   6 & 23.02 &   22 &  3.9 &  1.0 &    0.11 &      \\ 
ISO\_LHSS\_J105115+57431 & 10:51:15.150 & +57:43:19.05 & 1.9 &  0.62 $\pm$ 0.10 & 3 &   8 & 17.69 &  571 &  0.8 &  0.8 & $<$0.01 &      \\ 
ISO\_LHSS\_J104950+57202 & 10:49:50.720 & +57:20:20.66 & 1.8 &  0.62 $\pm$ 0.09 & 4 &  10 & 21.15 &   76 & -0.1 &  0.9 & $<$0.01 &      \\ 
ISO\_LHSS\_J104925+57030 & 10:49:25.627 & +57:03:03.60 & 2.0 &  0.62 $\pm$ 0.11 & 1 &   7 & 18.92 &  434 & -2.0 &  0.1 & $<$0.01 &      \\ 
ISO\_LHSS\_J105011+57345 & 10:50:11.261 & +57:34:56.62 & 1.9 &  0.62 $\pm$ 0.10 & 3 &   9 & 16.07* & 1357 &  1.2 & -0.7 & $<$0.01 & star \\ 
ISO\_LHSS\_J105417+57331 & 10:54:17.889 & +57:33:10.64 & 2.0 &  0.62 $\pm$ 0.11 & 3 &   7 & 21.60 &   55 & -0.9 &  1.6 &    0.01 &      \\ 
ISO\_LHSS\_J105104+57331 & 10:51:04.285 & +57:33:12.21 & 2.0 &  0.62 $\pm$ 0.11 & 3 &   8 & 21.31 &   72 &  1.7 &  0.5 &    0.01 &      \\ 
ISO\_LHSS\_J105505+57295 & 10:55:05.951 & +57:29:58.71 & 1.9 &  0.62 $\pm$ 0.10 & 3 &   9 & 24.17 &    5 & -1.2 & -0.1 &    0.03 &      \\ 
ISO\_LHSS\_J105218+57261 & 10:52:18.065 & +57:26:14.85 & 1.6 &  0.61 $\pm$ 0.08 & 6 &  15 & 18.62 &  724 &  0.9 & -1.6 & $<$0.01 &      \\ 
ISO\_LHSS\_J105339+57120 & 10:53:39.602 & +57:12:00.79 & 2.0 &  0.61 $\pm$ 0.11 & 2 &   7 &       &      &      &      &         &      \\ 
ISO\_LHSS\_J105429+57061 & 10:54:29.539 & +57:06:11.97 & 1.9 &  0.61 $\pm$ 0.10 & 2 &   8 & 23.95 &   40 &  0.7 &  1.1 &    0.03 &      \\ 
ISO\_LHSS\_J105400+57054 & 10:54:00.776 & +57:05:48.83 & 2.1 &  0.60 $\pm$ 0.12 & 3 &   6 & 21.68 &   24 & -2.3 &  4.3 &    0.06 &      \\ 
ISO\_LHSS\_J105258+57274 & 10:52:58.187 & +57:27:49.85 & 1.9 &  0.60 $\pm$ 0.10 & 4 &   9 & 19.39 &  418 & -1.3 &  1.0 & $<$0.01 &      \\ 
ISO\_LHSS\_J104954+57045 & 10:49:54.320 & +57:04:56.00 & 1.7 &  0.60 $\pm$ 0.09 & 3 &  12 & 20.23 &  157 &  0.8 & -0.1 & $<$0.01 &      \\ 
ISO\_LHSS\_J105003+57221 & 10:50:03.823 & +57:22:13.66 & 1.9 &  0.60 $\pm$ 0.09 & 2 &   9 & 20.77 &  102 &  0.3 &  0.3 & $<$0.01 &      \\ 
ISO\_LHSS\_J105054+57041 & 10:50:54.371 & +57:04:15.36 & 1.9 &  0.59 $\pm$ 0.09 & 2 &   9 & 24.97 &    3 & -0.9 &  1.4 &    0.08 &      \\ 
ISO\_LHSS\_J105428+57224 & 10:54:28.359 & +57:22:48.83 & 1.8 &  0.59 $\pm$ 0.09 & 3 &  10 & 18.42 &  679 & -1.8 & -1.0 & $<$0.01 &      \\ 
ISO\_LHSS\_J105354+57190 & 10:53:54.357 & +57:19:06.45 & 1.9 &  0.59 $\pm$ 0.10 & 3 &   9 & 23.53 &   11 &  0.4 & -2.5 &    0.07 &      \\ 
ISO\_LHSS\_J105046+57075 & 10:50:46.351 & +57:07:54.33 & 1.7 &  0.59 $\pm$ 0.09 & 4 &  11 & 21.28 &   62 &  0.8 & -0.6 & $<$0.01 &      \\ 
ISO\_LHSS\_J105426+57065 & 10:54:26.646 & +57:06:55.21 & 2.0 &  0.59 $\pm$ 0.10 & 2 &   8 & 22.45 &   18 & -1.5 & -1.4 &    0.02 &      \\ 
ISO\_LHSS\_J105233+57093 & 10:52:33.424 & +57:09:35.49 & 2.0 &  0.59 $\pm$ 0.10 & 3 &   7 & 19.25 &  339 & -0.1 &  0.1 & $<$0.01 &      \\ 
ISO\_LHSS\_J105034+57334 & 10:50:34.186 & +57:33:45.55 & 1.8 &  0.58 $\pm$ 0.09 & 4 &  10 & 20.64 &  157 &  0.5 & -0.2 & $<$0.01 & star \\ 
ISO\_LHSS\_J105028+57135 & 10:50:28.015 & +57:13:50.48 & 1.8 &  0.58 $\pm$ 0.09 & 4 &  11 & 21.09 &   93 &  0.3 & -1.3 & $<$0.01 &      \\ 
ISO\_LHSS\_J105315+57194 & 10:53:15.736 & +57:19:41.15 & 1.9 &  0.58 $\pm$ 0.09 & 3 &   9 & 20.73 &   78 &  2.2 &  0.2 &    0.01 &      \\ 
ISO\_LHSS\_J104941+57113 & 10:49:41.536 & +57:11:35.82 & 2.0 &  0.57 $\pm$ 0.10 & 1 &   7 & 23.51 &   12 & -0.5 & -2.9 &    0.09 &      \\ 
ISO\_LHSS\_J105041+57372 & 10:50:41.609 & +57:37:24.26 & 1.7 &  0.57 $\pm$ 0.08 & 4 &  11 & 20.47 &  101 & -1.2 &  0.3 & $<$0.01 &      \\ 
ISO\_LHSS\_J105112+57121 & 10:51:12.590 & +57:12:13.96 & 1.8 &  0.57 $\pm$ 0.09 & 5 &  10 & 23.44 &   11 & -0.5 & -0.0 & $<$0.01 &      \\ 
ISO\_LHSS\_J105000+57101 & 10:50:00.399 & +57:10:15.69 & 1.9 &  0.56 $\pm$ 0.09 & 2 &   9 & 21.02 &   73 &  1.3 &  0.4 & $<$0.01 &      \\ 
ISO\_LHSS\_J105112+57311 & 10:51:12.748 & +57:31:16.08 & 2.0 &  0.56 $\pm$ 0.10 & 2 &   7 & 21.52 &   60 &  0.0 & -1.5 & $<$0.01 &      \\ 
ISO\_LHSS\_J105157+57084 & 10:51:57.653 & +57:08:46.29 & 2.1 &  0.56 $\pm$ 0.10 & 3 &   7 & 19.62 &  293 & -0.1 & -1.2 & $<$0.01 &      \\ 
ISO\_LHSS\_J105022+57244 & 10:50:22.240 & +57:24:42.25 & 1.9 &  0.55 $\pm$ 0.09 & 3 &   9 & 22.13 &   52 &  0.1 &  0.1 & $<$0.01 &      \\ 
ISO\_LHSS\_J105133+57332 & 10:51:33.486 & +57:33:27.39 & 2.0 &  0.55 $\pm$ 0.10 & 2 &   7 & 23.42 &   16 & -0.3 &  2.7 &    0.07 &      \\ 
   \noalign{\smallskip}
   \hline
\end{tabular}
\end{table*}

\newpage

\setcounter{table}{2}
\begin{table*}
\caption[]{Continue.}
\label{tab:catalog1}
\tiny
\begin{tabular}{c ccc ccc cc rr cl}
   \hline
   \noalign{\smallskip}
   Source name &    RA   &    DEC  & $\Delta$ & Flux   &  N$_{fr}$ &SNR  & r'  &SNR\_r&\multicolumn{2}{c}{$\Delta$(ISO-opt)}& Prob & Notes\\
               & (J2000) & (J2000) & [arcsec] & [mJy]  &           &    &   [mag]&    &\multicolumn{2}{c}{ [arcsec] }        &      &      \\
   \noalign{\smallskip}
   \hline
   \noalign{\smallskip}
ISO\_LHSS\_J105004+57273 & 10:50:04.922 & +57:27:33.35 & 2.1 &  0.55 $\pm$ 0.10 & 2 &   7 & 23.18 &   14 &  0.0 & -1.9 &    0.03 &      \\ 
ISO\_LHSS\_J105421+57142 & 10:54:21.006 & +57:14:26.35 & 2.0 &  0.55 $\pm$ 0.10 & 4 &   7 & 20.37 &   30 & -0.1 &  0.9 & $<$0.01 &      \\ 
ISO\_LHSS\_J104944+57113 & 10:49:44.707 & +57:11:32.87 & 2.2 &  0.55 $\pm$ 0.12 & 2 &   5 & 24.17 &    7 &  0.2 & -1.8 &    0.06 &      \\ 
ISO\_LHSS\_J105011+57400 & 10:50:11.514 & +57:40:07.64 & 1.9 &  0.55 $\pm$ 0.09 & 3 &   9 & 23.98 &    6 & -0.1 &  2.5 &    0.09 &      \\ 
ISO\_LHSS\_J105053+57242 & 10:50:53.177 & +57:24:25.55 & 2.0 &  0.55 $\pm$ 0.10 & 2 &   7 & 19.05 &  603 & -1.8 & -0.5 & $<$0.01 &      \\ 
ISO\_LHSS\_J105058+57335 & 10:50:58.455 & +57:33:53.68 & 1.8 &  0.54 $\pm$ 0.08 & 3 &  11 & 20.92 &  126 &  0.3 & -1.9 & $<$0.01 & star \\ 
ISO\_LHSS\_J105228+57114 & 10:52:28.011 & +57:11:46.04 & 2.0 &  0.54 $\pm$ 0.09 & 3 &   8 & 18.37 &  517 & -1.1 & -2.5 & $<$0.01 &      \\ 
ISO\_LHSS\_J105142+57371 & 10:51:42.590 & +57:37:12.79 & 2.1 &  0.54 $\pm$ 0.10 & 3 &   6 & 19.00 &  236 &  0.1 & -1.8 & $<$0.01 &      \\ 
ISO\_LHSS\_J105253+57241 & 10:52:53.741 & +57:24:17.69 & 1.9 &  0.54 $\pm$ 0.09 & 4 &   9 & 21.65 &   63 &  0.4 & -0.2 & $<$0.01 &      \\ 
ISO\_LHSS\_J105033+57170 & 10:50:33.245 & +57:17:02.46 & 2.0 &  0.54 $\pm$ 0.09 & 3 &   8 & 22.19 &   30 &  0.6 & -1.3 &    0.01 &      \\ 
ISO\_LHSS\_J105434+57173 & 10:54:34.871 & +57:17:37.60 & 1.9 &  0.53 $\pm$ 0.09 & 4 &   8 & 23.39 &   14 & -1.6 &  1.4 &    0.04 &      \\ 
ISO\_LHSS\_J105246+57063 & 10:52:46.930 & +57:06:36.85 & 2.0 &  0.53 $\pm$ 0.09 & 3 &   8 & 14.88* &  452 &  1.8 &  0.1 & $<$0.01 & star \\ 
ISO\_LHSS\_J105255+57222 & 10:52:55.049 & +57:22:24.20 & 1.9 &  0.53 $\pm$ 0.09 & 4 &   9 & 18.83 &  362 & -0.3 &  1.3 & $<$0.01 &      \\ 
ISO\_LHSS\_J105142+57212 & 10:51:42.946 & +57:21:22.15 & 2.0 &  0.53 $\pm$ 0.09 & 3 &   7 & 18.65 &  494 & -1.6 & -1.3 & $<$0.01 &      \\ 
ISO\_LHSS\_J105344+56553 & 10:53:44.462 & +56:55:36.05 & 2.2 &  0.53 $\pm$ 0.11 & 1 &   6 &       &      &      &      &         &      \\ 
ISO\_LHSS\_J105141+57410 & 10:51:41.448 & +57:41:09.45 & 1.8 &  0.53 $\pm$ 0.08 & 4 &  10 &       &      &      &      &         &      \\ 
ISO\_LHSS\_J105128+57285 & 10:51:28.967 & +57:28:54.01 & 1.8 &  0.53 $\pm$ 0.08 & 3 &  10 & 20.74 &  162 &  0.0 &  0.3 & $<$0.01 &      \\ 
ISO\_LHSS\_J105058+57251 & 10:50:58.264 & +57:25:14.99 & 2.0 &  0.53 $\pm$ 0.09 & 2 &   8 & 11.88 & 2171 & -0.1 &  1.9 & $<$0.01 & star \\ 
ISO\_LHSS\_J104929+57035 & 10:49:29.813 & +57:03:51.50 & 2.1 &  0.52 $\pm$ 0.10 & 2 &   6 &       &      &      &      &         &      \\ 
ISO\_LHSS\_J104953+57082 & 10:49:53.126 & +57:08:29.26 & 2.0 &  0.52 $\pm$ 0.09 & 3 &   7 &       &      &      &      &         &      \\ 
ISO\_LHSS\_J105242+57313 & 10:52:42.964 & +57:31:39.73 & 2.0 &  0.52 $\pm$ 0.09 & 2 &   8 & 17.07 & 1551 &  0.6 &  1.9 & $<$0.01 &      \\ 
ISO\_LHSS\_J105159+57241 & 10:51:59.927 & +57:24:11.02 & 2.0 &  0.52 $\pm$ 0.09 & 3 &   8 & 25.26 &    3 &  3.9 & -0.1 &    0.40 &      \\ 
ISO\_LHSS\_J105455+57304 & 10:54:55.591 & +57:30:48.38 & 1.9 &  0.52 $\pm$ 0.09 & 4 &   8 & 20.90 &  113 & -1.1 & -0.1 & $<$0.01 &      \\ 
ISO\_LHSS\_J105025+57020 & 10:50:25.170 & +57:02:08.02 & 2.3 &  0.51 $\pm$ 0.12 & 2 &   5 & 22.82 &   17 & -2.2 &  3.3 &    0.09 &      \\ 
ISO\_LHSS\_J104958+57384 & 10:49:58.616 & +57:38:44.13 & 2.0 &  0.51 $\pm$ 0.09 & 2 &   7 &       &      &      &      &         &      \\ 
ISO\_LHSS\_J105113+57265 & 10:51:13.323 & +57:26:53.69 & 2.0 &  0.51 $\pm$ 0.09 & 4 &   7 & 23.11 &   23 & -1.1 & -0.4 &    0.01 &      \\ 
ISO\_LHSS\_J105035+57260 & 10:50:35.867 & +57:26:08.22 & 1.9 &  0.51 $\pm$ 0.09 & 3 &   8 & 18.67 &  639 &  1.2 &  0.8 & $<$0.01 &      \\ 
ISO\_LHSS\_J105109+57252 & 10:51:09.576 & +57:25:26.59 & 1.9 &  0.51 $\pm$ 0.08 & 3 &   9 & 21.09 &  107 &  0.6 &  2.5 &    0.01 &      \\ 
ISO\_LHSS\_J104941+57035 & 10:49:41.796 & +57:03:50.38 & 2.0 &  0.51 $\pm$ 0.09 & 3 &   7 & 21.71 &   18 &  1.9 &  0.0 &    0.01 &      \\ 
ISO\_LHSS\_J105231+57320 & 10:52:31.190 & +57:32:04.13 & 2.1 &  0.50 $\pm$ 0.09 & 3 &   6 & 22.75 &   25 & -0.3 & -1.1 &    0.01 & star \\ 
ISO\_LHSS\_J105446+57383 & 10:54:46.263 & +57:38:30.23 & 2.2 &  0.50 $\pm$ 0.11 & 2 &   5 &       &      &      &      &         &      \\ 
ISO\_LHSS\_J105006+57452 & 10:50:06.914 & +57:45:28.32 & 1.9 &  0.50 $\pm$ 0.08 & 3 &   8 & 21.73 &   46 & -1.7 &  0.2 &    0.01 &      \\ 
ISO\_LHSS\_J105229+57302 & 10:52:29.150 & +57:30:23.99 & 2.0 &  0.50 $\pm$ 0.09 & 4 &   7 & 20.66 &  165 & -0.5 &  0.7 & $<$0.01 & star \\ 
ISO\_LHSS\_J105235+57233 & 10:52:35.182 & +57:23:31.37 & 2.1 &  0.49 $\pm$ 0.10 & 3 &   6 & 19.95 &  302 & -0.0 &  1.9 & $<$0.01 &      \\ 
ISO\_LHSS\_J105245+57142 & 10:52:45.923 & +57:14:23.83 & 2.0 &  0.49 $\pm$ 0.08 & 4 &   8 & 21.15 &   98 &  0.1 & -0.5 & $<$0.01 &      \\ 
ISO\_LHSS\_J105148+57324 & 10:51:48.794 & +57:32:48.74 & 1.9 &  0.49 $\pm$ 0.08 & 4 &   9 & 22.26 &   22 &  0.6 & -0.1 & $<$0.01 &      \\ 
ISO\_LHSS\_J105023+57243 & 10:50:23.577 & +57:24:37.29 & 2.0 &  0.49 $\pm$ 0.08 & 3 &   8 & 23.50 &   15 &  1.8 & -1.7 &    0.06 &      \\ 
ISO\_LHSS\_J105308+57064 & 10:53:08.221 & +57:06:45.82 & 2.0 &  0.49 $\pm$ 0.09 & 3 &   7 & 22.07 &   27 & -0.5 & -0.1 & $<$0.01 &      \\ 
ISO\_LHSS\_J105056+57163 & 10:50:56.781 & +57:16:30.77 & 1.7 &  0.49 $\pm$ 0.07 & 6 &  11 & 20.80 &  121 & -1.2 &  1.1 & $<$0.01 &      \\ 
ISO\_LHSS\_J104957+57175 & 10:49:57.601 & +57:17:59.91 & 2.0 &  0.48 $\pm$ 0.09 & 3 &   7 & 20.20 &  178 &  1.2 &  0.6 & $<$0.01 &      \\ 
ISO\_LHSS\_J105151+57133 & 10:51:51.207 & +57:13:31.26 & 1.7 &  0.48 $\pm$ 0.07 & 7 &  12 & 18.83 &  418 & -1.0 & -1.3 & $<$0.01 &      \\ 
ISO\_LHSS\_J105228+57421 & 10:52:28.824 & +57:42:14.38 & 2.0 &  0.48 $\pm$ 0.09 & 2 &   7 & 17.12 & 1810 &  0.6 &  0.9 & $<$0.01 &      \\ 
ISO\_LHSS\_J105236+57085 & 10:52:36.705 & +57:08:50.97 & 2.0 &  0.48 $\pm$ 0.09 & 3 &   7 & 22.98 &   17 &  1.0 &  0.7 &    0.01 &      \\ 
ISO\_LHSS\_J105235+57090 & 10:52:35.647 & +57:09:05.54 & 2.0 &  0.48 $\pm$ 0.08 & 4 &   7 & 25.33 &      &      &      &         &      \\ 
ISO\_LHSS\_J105155+57190 & 10:51:55.964 & +57:19:09.93 & 2.1 &  0.48 $\pm$ 0.09 & 3 &   6 & 19.31 &  494 & -0.4 & -0.3 & $<$0.01 &      \\ 
ISO\_LHSS\_J104946+57320 & 10:49:46.132 & +57:32:06.63 & 2.0 &  0.47 $\pm$ 0.08 & 3 &   8 & 20.77 &  107 & -1.7 &  1.7 &    0.01 &      \\ 
ISO\_LHSS\_J105112+57172 & 10:51:12.887 & +57:17:24.29 & 2.3 &  0.47 $\pm$ 0.11 & 2 &   5 &       &      &      &      &         &      \\ 
ISO\_LHSS\_J105047+57401 & 10:50:47.421 & +57:40:17.88 & 1.9 &  0.47 $\pm$ 0.08 & 3 &   8 & 23.52 &   10 &  0.2 & -1.5 &    0.02 &      \\ 
ISO\_LHSS\_J105430+57220 & 10:54:30.322 & +57:22:09.13 & 2.0 &  0.47 $\pm$ 0.08 & 3 &   8 & 17.39 & 1357 & -0.5 & -0.9 & $<$0.01 &      \\ 
ISO\_LHSS\_J105249+57123 & 10:52:49.222 & +57:12:38.25 & 2.3 &  0.47 $\pm$ 0.11 & 3 &   5 & 24.60 &    3 &  0.2 & -0.0 & $<$0.01 &      \\ 
ISO\_LHSS\_J105251+57153 & 10:52:51.610 & +57:15:39.65 & 1.9 &  0.47 $\pm$ 0.08 & 3 &   8 & 23.36 &   17 &  2.8 &  1.4 &    0.09 &      \\ 
ISO\_LHSS\_J105355+57295 & 10:53:55.990 & +57:29:58.45 & 2.2 &  0.47 $\pm$ 0.09 & 3 &   6 & 18.37 &   54 & -1.2 & -0.6 & $<$0.01 &      \\ 
ISO\_LHSS\_J105100+57242 & 10:51:00.183 & +57:24:28.82 & 2.2 &  0.46 $\pm$ 0.09 & 3 &   6 & 20.89 &  155 & -4.2 & -1.6 &    0.02 &      \\ 
ISO\_LHSS\_J105328+57401 & 10:53:28.711 & +57:40:17.13 & 2.1 &  0.46 $\pm$ 0.09 & 2 &   6 & 24.11 &    3 &  1.2 &  1.0 &    0.04 &      \\ 
ISO\_LHSS\_J105305+57233 & 10:53:05.219 & +57:23:32.65 & 1.9 &  0.46 $\pm$ 0.08 & 3 &   8 & 21.67 &   52 & -0.3 &  2.2 &    0.01 &      \\ 
ISO\_LHSS\_J105219+57055 & 10:52:19.786 & +57:05:56.41 & 2.0 &  0.46 $\pm$ 0.08 & 4 &   8 & 23.17 &   16 &  0.1 &  1.0 &    0.01 &      \\ 
ISO\_LHSS\_J105442+57225 & 10:54:42.462 & +57:22:58.89 & 2.0 &  0.46 $\pm$ 0.08 & 4 &   7 & 18.08 &  639 & -0.3 &  1.0 & $<$0.01 &      \\ 
ISO\_LHSS\_J104923+57165 & 10:49:23.452 & +57:16:59.60 & 2.1 &  0.46 $\pm$ 0.08 & 3 &   7 & 22.97 &   15 & -0.1 & -0.9 &    0.01 &      \\ 
ISO\_LHSS\_J105019+57151 & 10:50:19.977 & +57:15:10.22 & 2.2 &  0.46 $\pm$ 0.09 & 2 &   5 & 19.88 &  213 & -1.1 & -1.1 & $<$0.01 &      \\ 
   \noalign{\smallskip}
   \hline
\end{tabular}
\begin{list}{}{}
\item[$^{\mathrm{*}}$] The source is saturated on the r' image.
\end{list}
\end{table*}

\renewcommand{\arraystretch}{1.0}
%


%% file: LHSStab4.tex
%

\renewcommand{\arraystretch}{0.9}
\begin{table*}
\caption[]{LW3 source catalog: faint detections}
\label{tab:catalog2}
\tiny
\begin{tabular}{c ccc ccc cc rr cl}
   \hline
   \noalign{\smallskip}
   Source name &    RA   &    DEC  & $\Delta$ & Flux   &  N$_{fr}$ &SNR  & r'  &SNR\_r&\multicolumn{2}{c}{$\Delta$(ISO-opt)}& Prob & Notes\\
               & (J2000) & (J2000) & [arcsec] & [mJy]  &           &    &   [mag]&    &\multicolumn{2}{c}{ [arcsec] }        &      &      \\
   \noalign{\smallskip}
   \hline
   \noalign{\smallskip}
ISO\_LHSS\_J105339+57310 & 10:53:39.639 & +57:31:03.35 & 2.0 &  0.45 $\pm$ 0.08 & 4 &   8 & 19.68 &  350 & -0.6 & -1.1 & $<$0.01 &      \\ 
ISO\_LHSS\_J105043+57172 & 10:50:43.297 & +57:17:29.61 & 2.2 &  0.45 $\pm$ 0.10 & 2 &   5 & 18.99 &  543 &  1.4 &  2.3 & $<$0.01 &      \\ 
ISO\_LHSS\_J105007+57280 & 10:50:07.661 & +57:28:04.44 & 2.2 &  0.45 $\pm$ 0.09 & 4 &   5 & 21.53 &   79 &  0.0 & -2.8 &    0.02 &      \\ 
ISO\_LHSS\_J104949+57165 & 10:49:49.017 & +57:16:57.18 & 2.2 &  0.45 $\pm$ 0.10 & 2 &   5 & 19.43 &  271 & -0.3 & -0.3 & $<$0.01 &      \\ 
ISO\_LHSS\_J105350+57294 & 10:53:50.823 & +57:29:47.20 & 2.2 &  0.45 $\pm$ 0.09 & 3 &   5 &       &      &      &      &         &      \\ 
ISO\_LHSS\_J105215+57131 & 10:52:15.487 & +57:13:18.06 & 2.0 &  0.44 $\pm$ 0.08 & 3 &   7 & 21.38 &   67 &  1.8 & -1.8 &    0.01 &      \\ 
ISO\_LHSS\_J105056+57321 & 10:50:56.818 & +57:32:18.13 & 1.8 &  0.44 $\pm$ 0.07 & 4 &  10 & 23.09 &   17 &  2.2 & -1.7 &    0.06 &      \\ 
ISO\_LHSS\_J105249+56595 & 10:52:49.175 & +56:59:57.29 & 1.9 &  0.44 $\pm$ 0.07 & 3 &   8 & 22.87 &   25 &  1.2 & -2.8 &    0.06 &      \\ 
ISO\_LHSS\_J105314+57030 & 10:53:14.048 & +57:03:09.46 & 2.2 &  0.44 $\pm$ 0.09 & 3 &   6 & 22.15 &   28 & -1.3 & -4.1 &    0.06 &      \\ 
ISO\_LHSS\_J105121+57055 & 10:51:21.738 & +57:05:57.65 & 2.0 &  0.44 $\pm$ 0.08 & 4 &   7 & 22.31 &   19 & -1.2 & -2.4 &    0.03 &      \\ 
ISO\_LHSS\_J105410+56594 & 10:54:10.294 & +56:59:49.87 & 2.0 &  0.44 $\pm$ 0.08 & 4 &   7 & 16.46 &  905 & -0.2 &  1.9 & $<$0.01 & star \\ 
ISO\_LHSS\_J105159+57015 & 10:51:59.268 & +57:01:57.02 & 2.0 &  0.44 $\pm$ 0.08 & 4 &   8 & 23.99 &    6 &  0.1 &  0.1 & $<$0.01 &      \\ 
ISO\_LHSS\_J105058+57254 & 10:50:58.865 & +57:25:44.92 & 2.1 &  0.44 $\pm$ 0.08 & 4 &   7 & 19.58 &  153 & -0.6 &  0.0 & $<$0.01 &      \\ 
ISO\_LHSS\_J105258+57092 & 10:52:58.246 & +57:09:24.70 & 2.1 &  0.43 $\pm$ 0.08 & 4 &   6 & 17.93 &  350 & -1.0 & -0.2 & $<$0.01 &      \\ 
ISO\_LHSS\_J105046+57105 & 10:50:46.516 & +57:10:52.29 & 2.0 &  0.43 $\pm$ 0.07 & 2 &   8 & 20.69 &  119 &  0.0 & -2.6 &    0.01 &      \\ 
ISO\_LHSS\_J105049+57252 & 10:50:49.255 & +57:25:28.12 & 2.2 &  0.43 $\pm$ 0.09 & 2 &   5 & 18.91 &  679 & -1.1 &  0.8 & $<$0.01 &      \\ 
ISO\_LHSS\_J105228+57294 & 10:52:28.136 & +57:29:47.23 & 2.1 &  0.43 $\pm$ 0.09 & 3 &   6 & 23.14 &   14 & -2.7 &  1.4 &    0.07 &  ?   \\ 
ISO\_LHSS\_J105328+57304 & 10:53:28.608 & +57:30:49.81 & 2.3 &  0.43 $\pm$ 0.10 & 2 &   5 &       &      &      &      &         &  ?   \\ 
ISO\_LHSS\_J105101+57034 & 10:51:01.285 & +57:03:42.32 & 1.9 &  0.43 $\pm$ 0.07 & 4 &   9 & 12.61 & 1551 & -1.6 &  0.3 & $<$0.01 & star \\ 
ISO\_LHSS\_J104938+57090 & 10:49:38.529 & +57:09:01.72 & 2.0 &  0.43 $\pm$ 0.07 & 3 &   8 & 22.05 &   38 & -0.7 & -1.7 &    0.01 &      \\ 
ISO\_LHSS\_J105308+57172 & 10:53:08.961 & +57:17:22.34 & 2.0 &  0.43 $\pm$ 0.08 & 4 &   7 & 20.85 &   63 &  0.1 &  1.1 & $<$0.01 &      \\ 
ISO\_LHSS\_J105137+57055 & 10:51:37.075 & +57:05:51.99 & 2.0 &  0.43 $\pm$ 0.07 & 3 &   8 & 21.06 &   85 & -0.8 & -2.4 &    0.01 &      \\ 
ISO\_LHSS\_J105258+57403 & 10:52:58.517 & +57:40:32.24 & 2.3 &  0.43 $\pm$ 0.10 & 3 &   5 & 23.29 &    6 &  0.0 &  0.5 & $<$0.01 &      \\ 
ISO\_LHSS\_J105201+57071 & 10:52:01.384 & +57:07:18.04 & 2.0 &  0.43 $\pm$ 0.07 & 6 &   8 & 19.48 &  170 &  1.1 & -1.0 & $<$0.01 &      \\ 
ISO\_LHSS\_J105218+57260 & 10:52:18.896 & +57:26:01.69 & 2.2 &  0.43 $\pm$ 0.09 & 3 &   5 & 23.05 &   47 & -2.5 & -3.0 &    0.11 &  ?   \\ 
ISO\_LHSS\_J104918+57145 & 10:49:18.245 & +57:14:56.35 & 2.3 &  0.43 $\pm$ 0.10 & 2 &   5 &       &      &      &      &         &      \\ 
ISO\_LHSS\_J105232+57215 & 10:52:32.457 & +57:21:57.70 & 2.1 &  0.43 $\pm$ 0.08 & 3 &   6 & 21.06 &  128 &  1.0 &  0.6 & $<$0.01 &      \\ 
ISO\_LHSS\_J105117+57142 & 10:51:17.834 & +57:14:24.65 & 2.0 &  0.42 $\pm$ 0.07 & 4 &   7 & 22.05 &   35 & -0.5 & -0.3 & $<$0.01 &      \\ 
ISO\_LHSS\_J104952+57263 & 10:49:52.628 & +57:26:36.59 & 1.9 &  0.42 $\pm$ 0.07 & 4 &   8 & 21.73 &      &  0.9 & -0.9 &         &      \\ 
ISO\_LHSS\_J105044+57172 & 10:50:44.718 & +57:17:24.68 & 2.1 &  0.42 $\pm$ 0.08 & 3 &   6 & 23.42 &   15 & -2.1 &  1.7 &    0.07 &      \\ 
ISO\_LHSS\_J105224+57224 & 10:52:24.979 & +57:22:45.86 & 1.9 &  0.42 $\pm$ 0.07 & 5 &   9 & 21.50 &  106 & -1.8 & -1.2 &    0.01 & star \\ 
ISO\_LHSS\_J105147+57085 & 10:51:47.190 & +57:08:55.08 & 2.0 &  0.42 $\pm$ 0.07 & 7 &   8 & 18.56 &  350 &  1.4 &  1.7 & $<$0.01 &      \\ 
ISO\_LHSS\_J105152+57244 & 10:51:52.108 & +57:24:47.29 & 2.2 &  0.41 $\pm$ 0.09 & 2 &   5 & 21.07 &  145 &  0.1 & -1.9 & $<$0.01 & star \\ 
ISO\_LHSS\_J105421+57380 & 10:54:21.262 & +57:38:03.84 & 2.1 &  0.41 $\pm$ 0.08 & 3 &   6 & 23.27 &   17 & -0.8 &  1.7 &    0.03 &      \\ 
ISO\_LHSS\_J104900+57071 & 10:49:00.289 & +57:07:16.39 & 2.1 &  0.41 $\pm$ 0.08 & 3 &   6 & 24.49 &    4 & -3.3 & -1.3 &    0.25 &      \\ 
ISO\_LHSS\_J105242+57125 & 10:52:42.550 & +57:12:51.43 & 2.3 &  0.41 $\pm$ 0.10 & 2 &   5 & 23.40 &   14 &  2.3 & -1.5 &    0.07 &      \\ 
ISO\_LHSS\_J105049+57031 & 10:50:49.274 & +57:03:13.70 & 2.3 &  0.41 $\pm$ 0.10 & 4 &   5 & 20.41 &   86 & -1.2 &  2.2 &    0.01 &      \\ 
ISO\_LHSS\_J105412+57121 & 10:54:12.184 & +57:12:10.30 & 2.0 &  0.41 $\pm$ 0.07 & 3 &   7 & 24.28 &    6 & -3.6 &  1.7 &    0.26 &      \\ 
ISO\_LHSS\_J105436+57202 & 10:54:36.793 & +57:20:27.74 & 2.2 &  0.41 $\pm$ 0.09 & 3 &   5 & 25.78 &    3 & -1.2 &  1.3 &    0.11 &      \\ 
ISO\_LHSS\_J105339+57010 & 10:53:39.463 & +57:01:04.68 & 2.2 &  0.41 $\pm$ 0.09 & 3 &   5 & 17.45 &  452 & -0.7 &  0.7 & $<$0.01 &      \\ 
ISO\_LHSS\_J105050+57084 & 10:50:50.087 & +57:08:45.19 & 2.3 &  0.41 $\pm$ 0.10 & 3 &   5 & 24.72 &    3 & -0.4 & -2.0 &    0.10 &      \\ 
ISO\_LHSS\_J105423+57194 & 10:54:23.331 & +57:19:48.43 & 2.0 &  0.41 $\pm$ 0.07 & 4 &   8 & 25.29 &    4 & -1.0 &  1.2 &    0.08 &      \\ 
ISO\_LHSS\_J105439+57382 & 10:54:39.474 & +57:38:24.49 & 2.2 &  0.41 $\pm$ 0.09 & 3 &   5 & 24.33 &    3 &  1.7 & -1.6 &    0.10 &      \\ 
ISO\_LHSS\_J105111+57331 & 10:51:11.906 & +57:33:12.55 & 2.1 &  0.40 $\pm$ 0.08 & 2 &   6 & 18.97 &  418 &  0.0 & -2.1 & $<$0.01 &      \\ 
ISO\_LHSS\_J105154+57343 & 10:51:54.529 & +57:34:37.43 & 2.1 &  0.40 $\pm$ 0.07 & 3 &   7 & 20.69 &  149 &  1.4 & -0.4 & $<$0.01 & star \\ 
ISO\_LHSS\_J105501+57361 & 10:55:01.729 & +57:36:17.26 & 2.2 &  0.40 $\pm$ 0.08 & 3 &   6 &       &      &      &      &         &      \\ 
ISO\_LHSS\_J105008+57032 & 10:50:08.577 & +57:03:22.64 & 2.2 &  0.40 $\pm$ 0.08 & 3 &   6 &       &      &      &      &         &      \\ 
ISO\_LHSS\_J105240+57305 & 10:52:40.085 & +57:30:56.61 & 2.2 &  0.40 $\pm$ 0.08 & 5 &   5 & 23.30 &   12 &  3.0 &  1.5 &    0.10 &  ?   \\ 
ISO\_LHSS\_J105459+57264 & 10:54:59.923 & +57:26:46.32 & 2.0 &  0.40 $\pm$ 0.07 & 3 &   7 & 23.83 &   13 & -0.2 & -2.0 &    0.06 &      \\ 
ISO\_LHSS\_J104952+57173 & 10:49:52.830 & +57:17:36.32 & 2.0 &  0.40 $\pm$ 0.07 & 4 &   8 & 22.77 &   16 &  2.2 & -1.5 &    0.04 &      \\ 
ISO\_LHSS\_J105049+57290 & 10:50:49.596 & +57:29:02.61 & 2.0 &  0.40 $\pm$ 0.07 & 3 &   7 &       &      &      &      &         &      \\ 
ISO\_LHSS\_J105006+57145 & 10:50:06.423 & +57:14:55.23 & 2.0 &  0.40 $\pm$ 0.07 & 3 &   7 & 23.20 &   11 & -1.0 & -2.9 &    0.08 &      \\ 
ISO\_LHSS\_J105454+57362 & 10:54:54.265 & +57:36:29.97 & 2.0 &  0.39 $\pm$ 0.07 & 4 &   7 &       &      &      &      &         &      \\ 
ISO\_LHSS\_J105308+56594 & 10:53:08.134 & +56:59:40.77 & 2.2 &  0.39 $\pm$ 0.08 & 3 &   5 & 21.15 &   59 & -1.4 & -0.6 & $<$0.01 &      \\ 
ISO\_LHSS\_J105252+57160 & 10:52:52.903 & +57:16:02.28 & 2.1 &  0.39 $\pm$ 0.08 & 4 &   6 & 20.62 &  170 & -0.9 & -1.0 & $<$0.01 &      \\ 
ISO\_LHSS\_J105223+57055 & 10:52:23.115 & +57:05:58.14 & 2.3 &  0.39 $\pm$ 0.09 & 3 &   5 & 23.73 &    8 &  0.0 & -1.7 &    0.04 &      \\ 
ISO\_LHSS\_J104920+57031 & 10:49:20.680 & +57:03:18.00 & 2.3 &  0.39 $\pm$ 0.09 & 1 &   5 & 24.83 &    4 & -1.4 & -0.2 &    0.06 &      \\ 
ISO\_LHSS\_J105447+57222 & 10:54:47.260 & +57:22:20.36 & 2.2 &  0.39 $\pm$ 0.08 & 3 &   6 &       &      &      &      &         &      \\ 
ISO\_LHSS\_J105125+57354 & 10:51:25.649 & +57:35:42.87 & 2.1 &  0.39 $\pm$ 0.07 & 3 &   7 & 15.62 & 2714 & -0.2 & -1.3 & $<$0.01 &      \\ 
ISO\_LHSS\_J105442+57340 & 10:54:42.393 & +57:34:03.99 & 2.1 &  0.39 $\pm$ 0.07 & 4 &   6 & 23.93 &    6 & -3.2 & -2.0 &    0.20 &      \\ 
ISO\_LHSS\_J105235+57063 & 10:52:35.819 & +57:06:32.10 & 2.1 &  0.39 $\pm$ 0.07 & 3 &   6 & 23.90 &    6 &  0.9 & -0.6 &    0.02 &      \\ 
ISO\_LHSS\_J105255+57115 & 10:52:55.668 & +57:11:51.12 & 2.2 &  0.38 $\pm$ 0.08 & 3 &   5 & 21.40 &   56 & -0.1 &  2.2 &    0.01 &      \\ 
ISO\_LHSS\_J105249+57272 & 10:52:49.900 & +57:27:23.64 & 2.3 &  0.38 $\pm$ 0.09 & 2 &   5 & 22.79 &    9 &  1.4 &  3.3 &    0.07 &      \\ 
ISO\_LHSS\_J105247+57362 & 10:52:47.743 & +57:36:20.10 & 1.9 &  0.38 $\pm$ 0.06 & 7 &   9 & 17.87 & 1357 & -0.9 & -0.5 & $<$0.01 &      \\ 
ISO\_LHSS\_J104927+57115 & 10:49:27.169 & +57:11:54.25 & 2.2 &  0.38 $\pm$ 0.08 & 3 &   5 & 25.02 &    3 &  0.4 &  2.0 &    0.12 &      \\ 
ISO\_LHSS\_J105033+57391 & 10:50:33.798 & +57:39:14.85 & 2.0 &  0.38 $\pm$ 0.07 & 4 &   7 & 22.23 &   39 & -2.3 &  1.2 &    0.02 & star \\ 
ISO\_LHSS\_J105329+56570 & 10:53:29.019 & +56:57:06.86 & 2.3 &  0.38 $\pm$ 0.09 & 2 &   5 & 24.11 &    6 &  2.6 &  1.3 &    0.13 &      \\ 
ISO\_LHSS\_J105442+57300 & 10:54:42.594 & +57:30:02.44 & 2.2 &  0.38 $\pm$ 0.08 & 3 &   5 & 23.29 &   12 &  0.3 &  1.1 &    0.01 &      \\ 
ISO\_LHSS\_J105159+57014 & 10:51:59.022 & +57:01:41.79 & 2.3 &  0.38 $\pm$ 0.08 & 3 &   5 &       &      &      &      &         &      \\ 
ISO\_LHSS\_J104958+57305 & 10:49:58.169 & +57:30:53.09 & 2.1 &  0.38 $\pm$ 0.07 & 3 &   6 & 24.91 &    4 & -1.0 & -1.2 &    0.07 &      \\ 
ISO\_LHSS\_J105226+57290 & 10:52:26.276 & +57:29:06.58 & 2.1 &  0.38 $\pm$ 0.07 & 3 &   6 &       &      &      &      &         &  ?   \\ 
ISO\_LHSS\_J105037+57304 & 10:50:37.042 & +57:30:40.00 & 2.0 &  0.38 $\pm$ 0.07 & 5 &   7 & 21.12 &  103 & -1.9 & -1.4 &    0.01 &      \\ 
ISO\_LHSS\_J104954+57031 & 10:49:54.023 & +57:03:12.89 & 2.2 &  0.38 $\pm$ 0.08 & 2 &   6 & 24.96 &    3 &  1.2 &  2.1 &    0.15 &      \\ 
ISO\_LHSS\_J105103+57093 & 10:51:03.472 & +57:09:36.88 & 2.0 &  0.37 $\pm$ 0.07 & 3 &   7 & 21.25 &   82 & -1.2 & -1.8 &    0.01 &      \\ 
ISO\_LHSS\_J105231+57155 & 10:52:31.395 & +57:15:50.34 & 2.1 &  0.37 $\pm$ 0.07 & 4 &   6 & 13.12 & 1085 &  0.4 & -3.4 & $<$0.01 & star \\ 
ISO\_LHSS\_J105041+57302 & 10:50:41.693 & +57:30:20.08 & 2.3 &  0.37 $\pm$ 0.09 & 3 &   5 & 21.83 &   54 &  2.0 & -1.4 &    0.02 &      \\ 
ISO\_LHSS\_J105240+57160 & 10:52:40.415 & +57:16:02.99 & 2.2 &  0.37 $\pm$ 0.08 & 3 &   5 & 22.42 &   35 &  0.6 &  3.2 &    0.05 &      \\ 
ISO\_LHSS\_J105302+57360 & 10:53:02.076 & +57:36:09.36 & 2.2 &  0.37 $\pm$ 0.08 & 3 &   6 & 22.88 &   17 & -1.2 & -3.5 &    0.08 &      \\ 
ISO\_LHSS\_J105126+57423 & 10:51:26.635 & +57:42:35.52 & 2.1 &  0.37 $\pm$ 0.07 & 4 &   6 & 14.77 & 3619 & -3.7 & -0.9 & $<$0.01 &      \\ 
ISO\_LHSS\_J104954+57313 & 10:49:54.379 & +57:31:39.55 & 2.3 &  0.37 $\pm$ 0.09 & 3 &   5 & 23.87 &    3 &  0.4 & -1.5 &    0.03 &      \\ 
ISO\_LHSS\_J105329+57182 & 10:53:29.696 & +57:18:22.60 & 2.2 &  0.37 $\pm$ 0.07 & 3 &   6 & 21.42 &   69 &  0.3 &  2.2 &    0.01 &      \\ 
ISO\_LHSS\_J105058+57185 & 10:50:58.315 & +57:18:59.04 & 2.3 &  0.37 $\pm$ 0.09 & 3 &   5 &       &      &      &      &         &  ?   \\ 
ISO\_LHSS\_J105336+57005 & 10:53:36.427 & +57:00:59.07 & 2.2 &  0.36 $\pm$ 0.07 & 4 &   5 & 22.16 &   28 &  3.3 &  1.5 &    0.05 &      \\ 
ISO\_LHSS\_J105218+57102 & 10:52:18.219 & +57:10:25.97 & 2.2 &  0.36 $\pm$ 0.07 & 4 &   6 & 21.48 &   54 &  0.0 & -0.2 & $<$0.01 &      \\ 
ISO\_LHSS\_J105008+57251 & 10:50:08.815 & +57:25:12.01 & 2.1 &  0.36 $\pm$ 0.07 & 5 &   6 &       &      &      &      &         &      \\ 
ISO\_LHSS\_J105309+57334 & 10:53:09.899 & +57:33:46.65 & 2.2 &  0.36 $\pm$ 0.07 & 4 &   5 & 23.07 &    8 & -1.1 &  2.3 &    0.05 &      \\ 
ISO\_LHSS\_J105050+57381 & 10:50:50.336 & +57:38:19.34 & 2.2 &  0.36 $\pm$ 0.07 & 3 &   6 & 19.04 &  603 &  1.6 & -0.8 & $<$0.01 & star \\ 
ISO\_LHSS\_J105236+57310 & 10:52:36.207 & +57:31:03.12 & 2.0 &  0.36 $\pm$ 0.06 & 3 &   7 & 21.04 &   91 & -0.6 & -0.9 & $<$0.01 &      \\ 
ISO\_LHSS\_J105220+57134 & 10:52:20.662 & +57:13:47.13 & 2.3 &  0.36 $\pm$ 0.08 & 3 &   5 & 24.89 &    4 &  0.2 & -1.7 &    0.08 &      \\ 
ISO\_LHSS\_J105314+57193 & 10:53:14.095 & +57:19:33.05 & 2.2 &  0.36 $\pm$ 0.08 & 3 &   5 & 22.09 &   11 & -2.7 & -1.1 &    0.03 &      \\ 
ISO\_LHSS\_J105249+57391 & 10:52:49.757 & +57:39:14.47 & 2.1 &  0.35 $\pm$ 0.07 & 4 &   6 &       &      &      &      &         &      \\ 
ISO\_LHSS\_J105104+57425 & 10:51:04.991 & +57:42:57.02 & 2.1 &  0.35 $\pm$ 0.07 & 4 &   6 &       &      &      &      &         &      \\ 
ISO\_LHSS\_J105104+57273 & 10:51:04.548 & +57:27:38.58 & 2.2 &  0.35 $\pm$ 0.07 & 4 &   6 & 20.49 &  217 &  0.7 &  1.5 & $<$0.01 &      \\ 
ISO\_LHSS\_J104953+57401 & 10:49:53.372 & +57:40:12.50 & 2.2 &  0.35 $\pm$ 0.07 & 3 &   6 & 22.89 &   17 & -1.8 &  0.6 &    0.02 &      \\ 
ISO\_LHSS\_J105348+57193 & 10:53:48.420 & +57:19:31.55 & 2.3 &  0.35 $\pm$ 0.08 & 3 &   5 &       &      &      &      &         &      \\ 
ISO\_LHSS\_J105110+57160 & 10:51:10.613 & +57:16:03.72 & 2.1 &  0.35 $\pm$ 0.06 & 3 &   7 & 20.78 &  113 & -1.5 &  1.3 & $<$0.01 &      \\ 
ISO\_LHSS\_J105016+57413 & 10:50:16.370 & +57:41:35.89 & 2.0 &  0.35 $\pm$ 0.06 & 4 &   7 & 23.64 &    7 & -3.5 & -0.1 &    0.14 &      \\ 
ISO\_LHSS\_J105150+57412 & 10:51:50.043 & +57:41:23.32 & 2.1 &  0.35 $\pm$ 0.07 & 3 &   6 & 20.11 &   61 &  0.0 & -1.8 & $<$0.01 &      \\ 
ISO\_LHSS\_J105027+57281 & 10:50:27.803 & +57:28:16.58 & 2.2 &  0.35 $\pm$ 0.08 & 3 &   5 & 19.84 & 1206 & -1.7 & -1.9 & $<$0.01 & star \\ 
   \noalign{\smallskip}
   \hline
\end{tabular}
\end{table*}
\newpage

\setcounter{table}{3}
\begin{table*}
\caption[]{Continue.}
\label{tab:catalog2}
\tiny
\begin{tabular}{c ccc ccc cc rr cl}
   \hline
   \noalign{\smallskip}
   Source name &    RA   &    DEC  & $\Delta$ & Flux   &  N$_{fr}$ &SNR  & r'  &SNR\_r&\multicolumn{2}{c}{$\Delta$(ISO-opt)}& Prob & Notes\\
               & (J2000) & (J2000) & [arcsec] & [mJy]  &           &    &   [mag]&    &\multicolumn{2}{c}{ [arcsec] }        &      &      \\
   \noalign{\smallskip}
   \hline
   \noalign{\smallskip}
ISO\_LHSS\_J105245+57074 & 10:52:45.476 & +57:07:47.55 & 2.1 &  0.35 $\pm$ 0.07 & 6 &   6 & 18.49 &  517 &  1.9 &  0.0 & $<$0.01 &      \\ 
ISO\_LHSS\_J105018+57431 & 10:50:18.618 & +57:43:19.24 & 2.2 &  0.35 $\pm$ 0.07 & 3 &   5 & 22.78 &   23 & -2.4 &  1.3 &    0.04 &      \\ 
ISO\_LHSS\_J105426+57072 & 10:54:26.649 & +57:07:26.17 & 2.2 &  0.34 $\pm$ 0.07 & 4 &   5 &       &      &      &      &         &      \\ 
ISO\_LHSS\_J105349+57264 & 10:53:49.127 & +57:26:44.91 & 2.2 &  0.34 $\pm$ 0.07 & 3 &   5 &       &      &      &      &         &      \\ 
ISO\_LHSS\_J105036+57054 & 10:50:36.548 & +57:05:45.84 & 2.2 &  0.34 $\pm$ 0.07 & 3 &   5 & 24.49 &    5 &  0.1 & -0.5 &    0.01 &      \\ 
ISO\_LHSS\_J105013+57403 & 10:50:13.085 & +57:40:38.95 & 2.1 &  0.34 $\pm$ 0.07 & 3 &   6 & 24.48 &    4 &  2.3 & -2.1 &    0.20 &      \\ 
ISO\_LHSS\_J105139+57224 & 10:51:39.690 & +57:22:48.84 & 2.0 &  0.34 $\pm$ 0.06 & 4 &   7 &       &      &      &      &         &  ?   \\ 
ISO\_LHSS\_J105038+57114 & 10:50:38.298 & +57:11:42.60 & 2.3 &  0.34 $\pm$ 0.08 & 4 &   5 & 24.45 &    5 &  0.5 &  0.2 &    0.01 &      \\ 
ISO\_LHSS\_J105039+57233 & 10:50:39.738 & +57:23:36.40 & 2.1 &  0.34 $\pm$ 0.07 & 4 &   6 & 19.26 &  724 &  1.5 & -0.3 & $<$0.01 & star \\ 
ISO\_LHSS\_J105400+57162 & 10:54:00.070 & +57:16:20.01 & 2.2 &  0.34 $\pm$ 0.07 & 3 &   6 & 23.18 &   11 &  0.2 & -1.6 &    0.02 &      \\ 
ISO\_LHSS\_J105237+57214 & 10:52:37.581 & +57:21:48.38 & 2.3 &  0.34 $\pm$ 0.08 & 3 &   5 & 21.01 &  123 &  1.7 & -0.0 & $<$0.01 &      \\ 
ISO\_LHSS\_J105033+57403 & 10:50:33.600 & +57:40:30.18 & 2.3 &  0.34 $\pm$ 0.08 & 2 &   5 & 24.12 &    3 & -3.7 & -0.6 &    0.22 &      \\ 
ISO\_LHSS\_J105300+57134 & 10:53:00.912 & +57:13:41.80 & 2.3 &  0.34 $\pm$ 0.08 & 3 &   5 &       &      &  2.7 & -0.6 & $<$0.01 & star \\ 
ISO\_LHSS\_J105104+57054 & 10:51:04.417 & +57:05:46.01 & 2.3 &  0.34 $\pm$ 0.08 & 2 &   5 & 22.59 &   26 &  2.9 &  3.1 &    0.09 &      \\ 
ISO\_LHSS\_J105303+56580 & 10:53:03.501 & +56:58:00.54 & 2.2 &  0.34 $\pm$ 0.07 & 3 &   5 & 22.48 &   20 &  0.4 & -0.3 & $<$0.01 &      \\ 
ISO\_LHSS\_J105052+57101 & 10:50:52.690 & +57:10:19.48 & 2.2 &  0.34 $\pm$ 0.07 & 4 &   6 & 23.72 &    9 & -0.5 &  0.6 &    0.01 &      \\ 
ISO\_LHSS\_J105006+57170 & 10:50:06.127 & +57:17:01.74 & 2.3 &  0.34 $\pm$ 0.08 & 4 &   5 & 20.39 &  119 &  0.2 & -0.1 & $<$0.01 &      \\ 
ISO\_LHSS\_J104932+57174 & 10:49:32.736 & +57:17:43.93 & 2.0 &  0.34 $\pm$ 0.06 & 3 &   7 & 20.72 &  139 &  1.0 & -1.6 & $<$0.01 &      \\ 
ISO\_LHSS\_J105311+57023 & 10:53:11.832 & +57:02:38.00 & 2.2 &  0.34 $\pm$ 0.07 & 4 &   5 & 23.20 &   14 & -0.0 & -2.5 &    0.05 &      \\ 
ISO\_LHSS\_J105047+57185 & 10:50:47.776 & +57:18:54.70 & 2.1 &  0.34 $\pm$ 0.07 & 3 &   6 & 22.74 &   33 & -0.4 &  0.1 & $<$0.01 &      \\ 
ISO\_LHSS\_J105117+57124 & 10:51:17.479 & +57:12:49.13 & 2.1 &  0.33 $\pm$ 0.07 & 3 &   6 &       &      &      &      &         &  ?   \\ 
ISO\_LHSS\_J104943+57110 & 10:49:43.945 & +57:11:07.19 & 2.2 &  0.33 $\pm$ 0.07 & 3 &   5 & 21.24 &   54 &  1.6 &  0.5 & $<$0.01 &      \\ 
ISO\_LHSS\_J105239+57312 & 10:52:39.016 & +57:31:24.36 & 2.3 &  0.33 $\pm$ 0.07 & 4 &   5 & 21.25 &   98 &  3.2 &  0.1 &    0.02 &      \\ 
ISO\_LHSS\_J105054+57164 & 10:50:54.749 & +57:16:40.66 & 2.1 &  0.33 $\pm$ 0.06 & 4 &   6 & 21.75 &   60 & -1.1 &  0.2 & $<$0.01 & star \\ 
ISO\_LHSS\_J105030+57055 & 10:50:30.249 & +57:05:56.59 & 2.2 &  0.33 $\pm$ 0.07 & 3 &   5 &       &      &      &      &         &      \\ 
ISO\_LHSS\_J105012+57034 & 10:50:12.935 & +57:03:48.38 & 2.2 &  0.33 $\pm$ 0.07 & 3 &   5 & 25.70 &    3 &  1.8 & -0.4 &    0.11 &      \\ 
ISO\_LHSS\_J105135+57255 & 10:51:35.713 & +57:25:52.22 & 2.2 &  0.33 $\pm$ 0.07 & 4 &   5 & 21.34 &  119 & -1.5 & -1.4 &    0.01 &      \\ 
ISO\_LHSS\_J105443+57355 & 10:54:43.557 & +57:35:54.30 & 2.3 &  0.33 $\pm$ 0.07 & 4 &   5 & 24.93 &    3 &  2.7 &  1.4 &    0.24 &      \\ 
ISO\_LHSS\_J105304+57093 & 10:53:04.270 & +57:09:32.51 & 2.2 &  0.32 $\pm$ 0.07 & 2 &   5 & 21.84 &   43 &  1.8 & -3.9 &    0.05 &      \\ 
ISO\_LHSS\_J105051+57292 & 10:50:51.013 & +57:29:23.54 & 2.1 &  0.32 $\pm$ 0.06 & 3 &   6 & 20.04 &  286 & -2.0 &  0.1 & $<$0.01 &      \\ 
ISO\_LHSS\_J105135+57251 & 10:51:35.966 & +57:25:12.54 & 2.1 &  0.32 $\pm$ 0.06 & 3 &   6 & 23.50 &   17 & -1.2 & -3.1 &    0.12 &      \\ 
ISO\_LHSS\_J105155+57163 & 10:51:55.313 & +57:16:34.44 & 2.0 &  0.32 $\pm$ 0.06 & 5 &   7 & 25.08 &    3 &  3.1 & -0.7 &    0.27 &  ?   \\ 
ISO\_LHSS\_J105147+57075 & 10:51:47.157 & +57:07:57.30 & 2.2 &  0.32 $\pm$ 0.06 & 7 &   6 & 21.96 &   31 &  1.0 & -1.8 &    0.01 &      \\ 
ISO\_LHSS\_J105010+57130 & 10:50:10.320 & +57:13:03.24 & 2.0 &  0.32 $\pm$ 0.06 & 5 &   7 & 20.19 &  145 & -0.2 &  2.2 & $<$0.01 &      \\ 
ISO\_LHSS\_J105154+57233 & 10:51:54.192 & +57:23:32.47 & 2.3 &  0.32 $\pm$ 0.07 & 4 &   5 & 20.93 &  149 &  1.6 & -0.7 & $<$0.01 &      \\ 
ISO\_LHSS\_J105029+57300 & 10:50:29.073 & +57:30:05.07 & 2.2 &  0.32 $\pm$ 0.07 & 3 &   5 & 21.32 &   79 & -1.7 & -1.8 &    0.01 &      \\ 
ISO\_LHSS\_J105219+57292 & 10:52:19.332 & +57:29:23.31 & 2.2 &  0.32 $\pm$ 0.06 & 6 &   6 & 21.70 &   62 &  0.7 &  0.8 & $<$0.01 &      \\ 
ISO\_LHSS\_J105340+57300 & 10:53:40.712 & +57:30:01.54 & 2.2 &  0.32 $\pm$ 0.07 & 5 &   5 & 23.06 &   19 & -0.1 &  1.4 &    0.01 &      \\ 
ISO\_LHSS\_J104931+57114 & 10:49:31.923 & +57:11:47.48 & 2.3 &  0.32 $\pm$ 0.07 & 3 &   5 &       &      &      &      &         &      \\ 
ISO\_LHSS\_J104926+57200 & 10:49:26.624 & +57:20:09.86 & 2.2 &  0.31 $\pm$ 0.07 & 4 &   5 & 21.98 &   32 & -0.7 &  0.4 & $<$0.01 &      \\ 
ISO\_LHSS\_J105059+57242 & 10:50:59.103 & +57:24:25.36 & 2.3 &  0.31 $\pm$ 0.07 & 3 &   5 & 12.61 & 1551 & -1.3 & -1.3 & $<$0.01 & star \\ 
ISO\_LHSS\_J105143+57290 & 10:51:43.989 & +57:29:05.86 & 2.1 &  0.31 $\pm$ 0.06 & 4 &   6 & 21.38 &  101 & -1.9 & -0.6 &    0.01 &      \\ 
ISO\_LHSS\_J105217+57405 & 10:52:17.417 & +57:40:50.56 & 1.9 &  0.31 $\pm$ 0.05 & 5 &   8 & 22.33 &   37 & -0.1 & -3.5 &    0.05 & star \\ 
ISO\_LHSS\_J105458+57284 & 10:54:58.920 & +57:28:44.74 & 2.2 &  0.31 $\pm$ 0.07 & 4 &   5 & 24.87 &    4 &  2.1 & -0.5 &    0.13 &      \\ 
ISO\_LHSS\_J105232+57244 & 10:52:32.029 & +57:24:47.75 & 2.2 &  0.31 $\pm$ 0.07 & 4 &   5 & 21.74 &   73 &  1.3 & -3.1 &    0.03 &      \\ 
ISO\_LHSS\_J105321+57131 & 10:53:21.742 & +57:13:14.37 & 2.3 &  0.31 $\pm$ 0.07 & 4 &   5 &       &      &      &      &         &      \\ 
ISO\_LHSS\_J105141+57160 & 10:51:41.136 & +57:16:03.91 & 2.3 &  0.31 $\pm$ 0.07 & 3 &   5 &       &      &      &      &         &      \\ 
ISO\_LHSS\_J105027+57440 & 10:50:27.209 & +57:44:04.44 & 2.2 &  0.31 $\pm$ 0.07 & 4 &   5 & 23.26 &   13 &  3.6 &  1.3 &    0.12 &      \\ 
ISO\_LHSS\_J105117+57163 & 10:51:17.692 & +57:16:39.87 & 2.0 &  0.31 $\pm$ 0.05 & 5 &   8 & 21.83 &   51 &  0.4 &  0.5 & $<$0.01 &  ?   \\ 
ISO\_LHSS\_J104945+57345 & 10:49:45.663 & +57:34:57.77 & 2.2 &  0.31 $\pm$ 0.06 & 3 &   6 & 24.91 &    4 &  0.3 & -3.5 &    0.30 &      \\ 
ISO\_LHSS\_J105330+57284 & 10:53:30.326 & +57:28:40.10 & 2.2 &  0.31 $\pm$ 0.07 & 4 &   5 &       &      &      &      &         &      \\ 
ISO\_LHSS\_J104955+57282 & 10:49:55.338 & +57:28:22.00 & 2.2 &  0.31 $\pm$ 0.07 & 4 &   5 & 22.72 &   19 &  1.3 & -1.6 &    0.02 &      \\ 
ISO\_LHSS\_J105141+57115 & 10:51:41.246 & +57:11:50.12 & 2.3 &  0.31 $\pm$ 0.07 & 4 &   5 & 21.03 &   72 & -0.1 &  0.5 & $<$0.01 &      \\ 
ISO\_LHSS\_J105315+57182 & 10:53:15.223 & +57:18:26.43 & 2.3 &  0.31 $\pm$ 0.07 & 3 &   5 & 21.08 &   34 &  0.4 & -1.5 & $<$0.01 &      \\ 
ISO\_LHSS\_J105318+57284 & 10:53:18.640 & +57:28:49.77 & 2.3 &  0.31 $\pm$ 0.07 & 6 &   5 & 20.48 &  101 & -0.1 & -1.5 & $<$0.01 &  ?   \\ 
ISO\_LHSS\_J105058+57182 & 10:50:58.019 & +57:18:25.42 & 2.2 &  0.31 $\pm$ 0.06 & 4 &   5 & 20.35 &  226 &  0.8 & -2.6 &    0.01 &      \\ 
ISO\_LHSS\_J105004+57085 & 10:50:04.545 & +57:08:51.64 & 2.2 &  0.30 $\pm$ 0.06 & 6 &   6 & 22.16 &   23 & -1.1 & -0.2 & $<$0.01 &      \\ 
ISO\_LHSS\_J105123+57122 & 10:51:23.317 & +57:12:26.59 & 2.2 &  0.30 $\pm$ 0.06 & 4 &   5 & 24.56 &    4 & -2.8 &  0.3 &    0.17 &      \\ 
ISO\_LHSS\_J105223+57085 & 10:52:23.727 & +57:08:55.27 & 2.2 &  0.30 $\pm$ 0.07 & 3 &   5 &       &      &      &      &         &      \\ 
ISO\_LHSS\_J104957+57275 & 10:49:57.524 & +57:27:58.23 & 2.3 &  0.30 $\pm$ 0.07 & 3 &   5 & 21.06 &   79 &  0.2 &  0.5 & $<$0.01 & star \\ 
ISO\_LHSS\_J105325+57074 & 10:53:25.052 & +57:07:41.36 & 2.2 &  0.30 $\pm$ 0.06 & 4 &   5 &       &      &      &      &         &      \\ 
ISO\_LHSS\_J104930+57115 & 10:49:30.857 & +57:11:57.05 & 2.3 &  0.30 $\pm$ 0.07 & 2 &   5 &       &      &      &      &         &      \\ 
ISO\_LHSS\_J105048+57022 & 10:50:48.461 & +57:02:22.99 & 2.2 &  0.30 $\pm$ 0.06 & 3 &   6 & 24.55 &    4 &  2.0 & -1.8 &    0.16 &      \\ 
ISO\_LHSS\_J105132+57111 & 10:51:32.571 & +57:11:11.14 & 2.2 &  0.30 $\pm$ 0.06 & 5 &   6 & 21.94 &   56 &  2.6 &  0.2 &    0.02 &      \\ 
ISO\_LHSS\_J105120+57300 & 10:51:20.065 & +57:30:02.31 & 2.1 &  0.30 $\pm$ 0.06 & 6 &   6 &       &      &      &      &         &      \\ 
ISO\_LHSS\_J104921+57100 & 10:49:21.910 & +57:10:09.80 & 2.1 &  0.30 $\pm$ 0.06 & 4 &   6 &       &      &      &      &         &      \\ 
ISO\_LHSS\_J105154+57133 & 10:51:54.056 & +57:13:30.15 & 2.3 &  0.30 $\pm$ 0.07 & 6 &   5 & 25.01 &    4 &  0.5 & -1.1 &    0.05 &      \\ 
ISO\_LHSS\_J105255+57110 & 10:52:55.005 & +57:11:06.05 & 2.2 &  0.29 $\pm$ 0.06 & 3 &   5 & 23.26 &   14 & -2.4 & -0.8 &    0.05 &  ?   \\ 
ISO\_LHSS\_J105101+57195 & 10:51:01.121 & +57:19:54.64 & 2.1 &  0.29 $\pm$ 0.06 & 5 &   6 & 21.89 &   49 &  0.6 &  1.4 &    0.01 &      \\ 
ISO\_LHSS\_J104953+57284 & 10:49:53.511 & +57:28:41.23 & 2.3 &  0.29 $\pm$ 0.07 & 4 &   5 & 22.39 &   31 &  0.9 & -0.4 & $<$0.01 &      \\ 
ISO\_LHSS\_J105350+57353 & 10:53:50.786 & +57:35:37.50 & 2.1 &  0.29 $\pm$ 0.06 & 6 &   6 & 23.49 &   12 &  1.7 &  3.3 &    0.14 &      \\ 
ISO\_LHSS\_J105245+57211 & 10:52:45.985 & +57:21:11.74 & 2.3 &  0.29 $\pm$ 0.07 & 4 &   5 &       &      &      &      &         &  ?   \\ 
ISO\_LHSS\_J105102+57145 & 10:51:02.439 & +57:14:59.63 & 2.3 &  0.29 $\pm$ 0.07 & 2 &   5 &       &      &      &      &         &      \\ 
ISO\_LHSS\_J105043+57062 & 10:50:43.612 & +57:06:27.89 & 2.2 &  0.29 $\pm$ 0.06 & 3 &   5 & 24.19 &    3 &  1.8 &  2.3 &    0.15 &      \\ 
ISO\_LHSS\_J105024+57142 & 10:50:24.507 & +57:14:24.39 & 2.3 &  0.29 $\pm$ 0.07 & 3 &   5 & 21.63 &   47 &  2.9 &  0.8 &    0.02 &      \\ 
ISO\_LHSS\_J105502+57322 & 10:55:02.677 & +57:32:25.28 & 2.3 &  0.29 $\pm$ 0.07 & 4 &   5 & 20.23 &   97 &  3.4 & -1.0 &    0.01 &      \\ 
ISO\_LHSS\_J105101+57095 & 10:51:01.630 & +57:09:53.84 & 2.3 &  0.29 $\pm$ 0.06 & 3 &   5 & 24.35 &    6 & -0.8 & -3.1 &    0.19 &      \\ 
ISO\_LHSS\_J105129+57163 & 10:51:29.656 & +57:16:39.92 & 2.2 &  0.29 $\pm$ 0.06 & 4 &   5 & 23.58 &   11 &  1.9 & -0.3 &    0.04 &      \\ 
ISO\_LHSS\_J105234+57182 & 10:52:34.156 & +57:18:26.56 & 2.3 &  0.29 $\pm$ 0.07 & 4 &   5 & 22.24 &   49 &  1.2 &  0.4 &    0.01 &  ?   \\ 
ISO\_LHSS\_J105424+57301 & 10:54:24.620 & +57:30:12.48 & 2.0 &  0.29 $\pm$ 0.05 & 6 &   7 & 21.43 &   44 & -2.2 &  1.1 &    0.01 &      \\ 
ISO\_LHSS\_J104913+57130 & 10:49:13.477 & +57:13:00.61 & 2.3 &  0.29 $\pm$ 0.06 & 3 &   5 & 22.10 &   30 &  1.4 & -0.5 &    0.01 &      \\ 
ISO\_LHSS\_J105306+57341 & 10:53:06.555 & +57:34:16.01 & 2.3 &  0.28 $\pm$ 0.06 & 4 &   5 &       &      &      &      &         &      \\ 
ISO\_LHSS\_J105220+57314 & 10:52:20.610 & +57:31:42.53 & 2.2 &  0.28 $\pm$ 0.06 & 7 &   5 & 24.34 &    6 & -0.4 & -1.5 &    0.05 &  ?   \\ 
ISO\_LHSS\_J105053+57285 & 10:50:53.558 & +57:28:55.57 & 2.3 &  0.28 $\pm$ 0.06 & 3 &   5 & 23.90 &   10 & -1.2 & -2.8 &    0.13 &      \\ 
ISO\_LHSS\_J104954+57244 & 10:49:54.236 & +57:24:45.57 & 2.3 &  0.28 $\pm$ 0.06 & 4 &   5 & 21.42 &   75 & -0.2 &  4.1 &    0.03 &      \\ 
ISO\_LHSS\_J105217+57034 & 10:52:17.021 & +57:03:49.79 & 2.3 &  0.27 $\pm$ 0.06 & 6 &   5 & 24.40 &    6 & -2.1 & -0.4 &    0.09 &      \\ 
ISO\_LHSS\_J105106+57042 & 10:51:06.368 & +57:04:25.62 & 2.3 &  0.27 $\pm$ 0.06 & 4 &   5 & 22.62 &   21 & -0.5 &  1.3 &    0.01 &      \\ 
ISO\_LHSS\_J105114+57434 & 10:51:14.608 & +57:43:47.42 & 2.3 &  0.27 $\pm$ 0.06 & 3 &   5 & 21.44 &   64 &  0.1 & -1.5 & $<$0.01 &      \\ 
ISO\_LHSS\_J104934+57315 & 10:49:34.607 & +57:31:51.06 & 2.3 &  0.27 $\pm$ 0.06 & 3 &   5 &       &      &      &      &         &      \\ 
ISO\_LHSS\_J105004+57372 & 10:50:04.080 & +57:37:25.15 & 2.2 &  0.27 $\pm$ 0.06 & 6 &   5 & 21.64 &   36 & -2.1 & -1.1 &    0.01 &      \\ 
ISO\_LHSS\_J105003+57355 & 10:50:03.607 & +57:35:56.30 & 2.2 &  0.27 $\pm$ 0.06 & 4 &   5 & 23.65 &    6 & -0.9 & -1.4 &    0.03 &      \\ 
ISO\_LHSS\_J105225+57313 & 10:52:25.876 & +57:31:30.61 & 2.2 &  0.27 $\pm$ 0.05 & 6 &   6 & 24.59 &    3 &  1.4 & -2.1 &    0.14 &      \\ 
ISO\_LHSS\_J104929+57072 & 10:49:29.473 & +57:07:25.89 & 2.3 &  0.27 $\pm$ 0.06 & 3 &   5 & 19.22 &  293 & -1.4 &  0.5 & $<$0.01 &      \\ 
ISO\_LHSS\_J105055+57160 & 10:50:55.723 & +57:16:06.44 & 2.3 &  0.27 $\pm$ 0.06 & 5 &   5 & 20.65 &  103 & -0.2 & -0.5 & $<$0.01 &      \\ 
ISO\_LHSS\_J105121+57193 & 10:51:21.584 & +57:19:37.83 & 2.3 &  0.27 $\pm$ 0.06 & 5 &   5 & 25.09 &    3 &  0.7 & -1.1 &    0.05 &      \\ 
ISO\_LHSS\_J105208+57320 & 10:52:08.335 & +57:32:05.29 & 2.2 &  0.27 $\pm$ 0.06 & 3 &   5 & 23.89 &   10 & -0.1 & -0.6 &    0.01 &      \\ 
ISO\_LHSS\_J104943+57240 & 10:49:43.209 & +57:24:00.14 & 2.1 &  0.26 $\pm$ 0.05 & 6 &   6 &       &      &      &      &         &      \\ 
ISO\_LHSS\_J105212+57364 & 10:52:12.349 & +57:36:47.19 & 2.3 &  0.26 $\pm$ 0.06 & 6 &   5 &       &      &      &      &         &      \\ 
   \noalign{\smallskip}
   \hline
\end{tabular}
\end{table*}
\renewcommand{\arraystretch}{1.0}

%
